\input harvmac
\input tables

\overfullrule=0pt
 
\def\A{{\scriptscriptstyle A}}
\def\B{{\scriptscriptstyle B}}
\def\C{{\scriptscriptstyle C}}
\def\D{{\scriptscriptstyle D}}

\def\I{{\scriptscriptstyle I}}
\def\J{{\scriptscriptstyle J}}
\def\K{{\scriptscriptstyle K}}
\def\L{{\scriptscriptstyle L}}

\def\Q{{\scriptscriptstyle Q}}
\def\R{{\scriptscriptstyle R}}
\def\S{{\scriptscriptstyle S}}
\def\T{{\scriptscriptstyle T}}

\def\X{{\scriptscriptstyle X}}
\def\Y{{\scriptscriptstyle Y}}
\def\Z{{\scriptscriptstyle Z}}
 

\def\CN{{\cal N}}

 
\def\a{\alpha}
\def\b{\beta}
\def\d{\delta}
\def\e{\epsilon}

\def\s{\sigma}
\def\t{\tau}


\def\half{{1 \over 2}}


\def\adot{{\dot{\alpha}}}
\def\bdot{{\dot{\beta}}}
\def\cdot{{\dot{\gamma}}}
\def\ddot{{\dot{\delta}}}

\def\bar#1{\overline{#1}}

\def\ccdot{\hbox{\kern-.1em$\cdot$\kern-.1em}}

\def\exyz{\e_{\rm \X \Y \Z}}
\def\Gdual{{\widetilde G}}
\def\glob{{\rm global}}
\def\gtap{\raise.3ex\hbox{$>$\kern-.75em\lower1ex\hbox{$\sim$}}}

\def\imax{{i_{\rm max}}}
\def\L{{\scriptscriptstyle L}}

\def\LambdaL{{\Lambda_\L}}
\def\LambdaR{{\Lambda_\R}}

\def\local{{\rm local}}
\def\ltap{\raise.3ex\hbox{$<$\kern-.75em\lower1ex\hbox{$\sim$}}}
\def\Nc{N_c}
\def\Ncdual{{\tilde N_c}}
\def\Ncpdual{{{{{\tilde N_c}'}}}}
\def\Nf{{N_f}}

\def\NQ{{N_\Q}}
\def\Pf{{\rm Pf}\>}
\def\R{{\scriptscriptstyle R}}
\def\qbar{{\bar{q}}}

\def\qp{{q'}}

\def\Rtwo{ {\Nf-4 \over \Nf+4} }

\def\sp{\>\>}

\def\therefore{{\hbox{..}\kern-.43em \raise.5ex \hbox{.}}\>\>}

\def\Vslash{V\hskip-0.75 em / \hskip+0.30 em}
\def\wtilde{{\widetilde w}}
\def\Wdyn{W_{\rm dyn}}
\def\WdynL{W^\L_{\rm dyn}}
\def\WdynR{W^\R_{\rm dyn}}

\def\Wmag{W_{\rm mag}}
\def\Wtree{W_{\rm tree}}
\def\xrm{{\rm \X}}
\def\yrm{{\rm \Y}}
\def\zrm{{\rm \Z}}

\newdimen\pmboffset
\pmboffset 0.022em
\def\oldpmb#1{\setbox0=\hbox{#1}%
 \copy0\kern-\wd0
 \kern\pmboffset\raise 1.732\pmboffset\copy0\kern-\wd0
 \kern\pmboffset\box0}
\def\pmb#1{\mathchoice{\oldpmb{$\displaystyle#1$}}{\oldpmb{$\textstyle#1$}}
      {\oldpmb{$\scriptstyle#1$}}{\oldpmb{$\scriptscriptstyle#1$}}}


\def\fund{  \> {\vcenter  {\vbox  
              {\hrule height.6pt
               \hbox {\vrule width.6pt  height5pt  
                      \kern5pt 
                      \vrule width.6pt  height5pt }
               \hrule height.6pt}
                         }
                   }
           \>\> }

\def\antifund{  \> \overline{ {\vcenter  {\vbox  
              {\hrule height.6pt
               \hbox {\vrule width.6pt  height5pt  
                      \kern5pt 
                      \vrule width.6pt  height5pt }
               \hrule height.6pt}
                         }
                   } }
           \>\> }

\def\sym{  \> {\vcenter  {\vbox  
              {\hrule height.6pt
               \hbox {\vrule width.6pt  height5pt  
                      \kern5pt 
                      \vrule width.6pt  height5pt 
                      \kern5pt
                      \vrule width.6pt height5pt}
               \hrule height.6pt}
                         }
              }
           \>\> }

\def\symbar{  \> \overline{ {\vcenter  {\vbox  
              {\hrule height.6pt
               \hbox {\vrule width.6pt  height5pt  
                      \kern5pt 
                      \vrule width.6pt  height5pt 
                      \kern5pt
                      \vrule width.6pt height5pt}
               \hrule height.6pt}
                         }
              }
           } \>\> }

\def\anti{ \> {\vcenter  {\vbox  
              {\hrule height.6pt
               \hbox {\vrule width.6pt  height5pt  
                      \kern5pt 
                      \vrule width.6pt  height5pt }
               \hrule height.6pt
               \hbox {\vrule width.6pt  height5pt  
                      \kern5pt 
                      \vrule width.6pt  height5pt }
               \hrule height.6pt}
                         }
              }
           \>\> }

\def\antithree{ \> 
              {\vcenter  {\vbox  
              {\hrule height.6pt
               \hbox {\vrule width.6pt  height5pt  
                      \kern5pt 
                      \vrule width.6pt  height5pt }
               \hrule height.6pt
               \hbox {\vrule width.6pt  height5pt  
                      \kern5pt 
                      \vrule width.6pt  height5pt }
               \hrule height.6pt
               \hbox {\vrule width.6pt  height5pt  
                      \kern5pt 
                      \vrule width.6pt  height5pt }
               \hrule height.6pt}
                         }
              }
           \>\> }

\def\antifour{ \> 
              {\vcenter  {\vbox  
              {\hrule height.6pt
               \hbox {\vrule width.6pt  height5pt  
                      \kern5pt 
                      \vrule width.6pt  height5pt }
               \hrule height.6pt
               \hbox {\vrule width.6pt  height5pt  
                      \kern5pt 
                      \vrule width.6pt  height5pt }
               \hrule height.6pt
               \hbox {\vrule width.6pt  height5pt  
                      \kern5pt 
                      \vrule width.6pt  height5pt }
               \hrule height.6pt
               \hbox {\vrule width.6pt  height5pt  
                      \kern5pt 
                      \vrule width.6pt  height5pt }
               \hrule height.6pt}
                         }
              }
           \>\> }

\def\antifive{ \> 
              {\vcenter  {\vbox  
              {\hrule height.6pt
               \hbox {\vrule width.6pt  height5pt  
                      \kern5pt 
                      \vrule width.6pt  height5pt }
               \hrule height.6pt
               \hbox {\vrule width.6pt  height5pt  
                      \kern5pt 
                      \vrule width.6pt  height5pt }
               \hrule height.6pt
               \hbox {\vrule width.6pt  height5pt  
                      \kern5pt 
                      \vrule width.6pt  height5pt }
               \hrule height.6pt
               \hbox {\vrule width.6pt  height5pt  
                      \kern5pt 
                      \vrule width.6pt  height5pt }
               \hrule height.6pt
               \hbox {\vrule width.6pt  height5pt  
                      \kern5pt 
                      \vrule width.6pt  height5pt }
               \hrule height.6pt}
                         }
              }
           \>\> }

\def\twotwo{
              {\vcenter  {\vbox
              {\hrule height.5pt
               \hbox {\vrule width.5pt  height4pt
                      \kern4pt
                      \vrule width.5pt  height4pt
                      \kern4pt
                      \vrule width.5pt height4pt}
               \hrule height.5pt
               \hbox {\vrule width.5pt  height4pt
                      \kern4pt
                      \vrule width.5pt  height4pt
                      \kern4pt
                      \vrule width.5pt height4pt}
               \hrule height.5pt}
                         }
              }
           \>\> }

\def\four{  {\vcenter  {\vbox
              {\hrule height.5pt
               \hbox {\vrule width.5pt  height4pt
                      \kern4pt
                      \vrule width.5pt  height4pt
                      \kern4pt
                      \vrule width.5pt  height4pt
                      \kern4pt
                      \vrule width.5pt  height4pt
                      \kern4pt
                      \vrule width.5pt height4pt}
               \hrule height.5pt}
                         }
              }
           }


\nref\SeibergI{N. Seiberg, Nucl. Phys. {\bf B435} (1995) 129.}
\nref\Pouliot{P. Pouliot, Phys. Lett. {\bf B359} (1995) 108.}
\nref\PouliotStrasslerII{P. Pouliot and M. Strassler, Phys. Lett. {\bf B375} 
 (1996) 175.}
\nref\IntriligatorThomas{K. Intriligator and S. Thomas, Nucl. Phys. 
  {\bf B473} (1996) 121; hep-th 9608046.}
\nref\PouliotII{P. Pouliot, Phys. Lett. {\bf B367} (1996) 151.}
\nref\PST{E. Poppitz, Y. Shadmi and S. Trivedi, Nucl. Phys. {\bf B480}
  (1996) 125; Phys. Lett. {\bf B388} (1996) 561.}
\nref\SSBMIT{
 C. Csaki, L. Randall and W. Skiba, Nucl. Phys. {\bf B479} (1996) 65; 
 C. Csaki, L. Randall, W. Skiba and R.G. Leigh, Phys. Lett. {\bf B387} (1996) 
  791; 
 L. Randall, hep-ph 9612426;
 W. Skiba, hep-th 9703159;
 R.G. Leigh, L. Randall and R. Rattazzi, hep-ph/9704246.}
\nref\Hotta{T. Hotta, K.-I. Izawa and T. Yanagida, Phys. Rev. {\bf D55} (1997) 
  415.}
\nref\Chou{C.-L. Chou, Phys. Lett. {\bf B391} (1997) 329.}
\nref\Bershadsky{M. Bershadsky, A. Johansen, T. Pantev, V. Sadov and C. Vafa,
  hep-th 9612052.}
\nref\Vafa{C. Vafa and B. Zweibach, hep-th 9701015.}
\nref\OoguriVafa{H. Ooguri and C. Vafa, hep-th 9702180.}
\nref\ElitzurI{S. Elitzur, A. Giveon and D. Kutasov, hep-th 9702014.}
\nref\ElitzurII{S. Elitzur, A. Giveon, D. Kutasov, E. Rabinovici and A. 
  Schwimmer, hep-th 9704104.}
\nref\Brandhuber{A. Brandhuber, J. Sonnenschein, S. Theisen and 
  S. Yankielowicz, hep-th 9704044.}
\nref\Evans{N. Evans, C. Johnson and A.D. Shapere, hep-th 9703210.}
\nref\Barbon{J.L.F. Barbon, hep-th 9703051.}
\nref\BrodieHanany{J.H. Brodie and A. Hanany, hep-th 9704043.}
\nref\AharonyHanany{O. Aharony and A. Hanany, hep-th 9704170.}
\nref\IntriligatorSeiberg{K. Intriligator and N. Seiberg, Nucl. Phys. 
  {\bf B444} (1995) 125.}
\nref\Berkooz{M. Berkooz, Nucl. Phys. {\bf B452} (1995) 513.}
\nref\LST{M. Luty, M. Schmaltz and J. Terning, Phys. Rev. {\bf D54} (1996) 
  7815.}
\nref\Sakai{T. Sakai, hep-th 9701155.}
\nref\ILS{K. Intriligator, R.G. Leigh and M.J. Strassler, Nucl. Phys. 
  {\bf B456} (1995) 567.}
\nref\Brodie{J. Brodie, Nucl. Phys. {\bf B478} (1996) 123; 
  J.H. Brodie and M.J. Strassler, hep-th 9611197.}
\nref\Kutasov{D. Kutasov, Phys. Lett. {\bf B351} (1995) 230.}
\nref\PouliotStrasslerI{P. Pouliot and M. Strassler, Phys. Lett. {\bf B370} 
 (1996) 76.}
\nref\Kawano{T. Kawano, Prog. Theor. Phys. {\bf 95} (1996) 963.}
\nref\DistlerKarch{J. Distler and A. Karch, hep-th 9611088.}
\nref\ChoI{P. Cho, hep-th 9702059.}
\nref\Shifman{M.A. Shifman and A.I. Vainshtein, Nucl. Phys. {\bf B359} (1991) 
  571.}
\nref\Elashvili{A.G. Elashvili, Funk. Anal. Pril. {\bf 6} (1972) 51.}
\nref\ChoII{P. Cho, hep-th 9701020.}
\nref\Harvey{J. Harvey, D.B. Reiss and P. Ramond, Nucl. Phys. {\bf B199}
 (1982) 223.}
\nref\SeibergII{N. Seiberg, Phys. Rev. {\bf D49} (1994) 6857.}
\nref\CSS{C. Cs\'aki, M. Schmaltz and W. Skiba, hep-th 9612207.}
\nref\KutasovSchwimmer{D. Kutasov and A. Schwimmer, Phys. Lett. {\bf B354} 
  (1995) 315.}
\nref\Intriligator{K. Intriligator, Nucl. Phys. {\bf B448} (1995) 187.}
\nref\LS{R.G. Leigh and M.J. Strassler, Nucl. Phys. {\bf B447} (1995) 95.}
\nref\KSS{D. Kutasov, A. Schwimmer and N. Seiberg, Nucl. Phys. {\bf B459} 
  (1996) 455.}
\nref\Witten{E. Witten, Phys. Lett. {\bf 117B} (1982) 324.}
\nref\ADSI{I. Affleck, M. Dine and N. Seiberg, Phys. Rev. Lett. {\bf 51} 
  (1983) 1026.}
\nref\LSII{R.G. Leigh and M.J. Strassler, hep-th 9611020.}
\nref\Kennedy{A.D. Kennedy, J. Math. Phys. {\bf 22} (1981) 1330.}
\nref\ADSII{I. Affleck, M. Dine and N. Seiberg, Phys. Lett. {\bf 140B} (1984) 
  59.}
\nref\IntriligatorPouliot{K. Intriligator and P. Pouliot, Phys. Lett. 
  {\bf B353} (1995) 471.}
\nref\MJSSCGT{M.J. Strassler, in preparation.}
\nref\Maekawa{N. Maekawa and J. Sato, Prog. Theory. Phys. {\bf 96} (1996) 979.}
\nref\Wess{J. Wess, {\it Lecture Notes in Physics 77}, (Springer), 1978.}
\nref\West{P. West, {\it Introduction to Supersymmetry and Supergravity}, 2nd 
  ed., (World Scientific), 1990.}
\nref\WessBagger{P. Wess and J. Bagger, {\it Supersymmetry and Supergravity},
  (Princeton University Press) 1992.}
\nref\BerkoozII{M. Berkooz, Nucl. Phys. {\bf B466} (1996), 75.}

 
\nfig\flowdiag{Renormalization group flows along various flat directions of 
the $SO(10)$ model with $\Nf$ vectors and $\NQ$ spinors.  Non-singlet matter 
contents for each sub-theory are listed.  The flat direction which connects 
the $SO(7)$ model with $\Nf-3$ vectors and $2\NQ$ spinors to the $G_2$ theory 
is not displayed for clarity's sake.}


\def\LongTitle#1#2#3#4#5{\nopagenumbers\abstractfont
\hsize=\hstitle\rightline{#1}
\hsize=\hstitle\rightline{#2}
\hsize=\hstitle\rightline{#3}
\vskip 0.5in\centerline{\titlefont #4} \centerline{\titlefont #5}
\abstractfont\vskip .3in\pageno=0}
 
\LongTitle{HUTP-97/A014}{CALT 68-2109}{RU-97-24}
{Dual Descriptions of SO(10) SUSY Gauge Theories}
{with Arbitrary Numbers of Spinors and Vectors}

\centerline{
  Micha Berkooz${}^1$, Peter Cho${}^2$, Per Kraus${}^3$ and Matthew J.
  Strassler${}^{2,4}$}
\bigskip
\centerline{${}^1$ Department of Physics and Astronomy, Rutgers University, 
  Piscataway, NJ  08855}
\centerline{${}^2$ Lyman Laboratory, Harvard University, Cambridge, MA  02138}
\centerline{${}^3$ Lauritsen Laboratory, California Institute of Technology,
  Pasadena, CA  91125}
\centerline{${}^4$ School of Natural Sciences, Institute for Advanced Study,
  Princeton, NJ  08540}

\vskip 0.3in
\centerline{\bf Abstract}
\bigskip

	We examine the low energy structure of $\CN=1$ supersymmetric $SO(10)$ 
gauge theory with matter chiral superfields in $\NQ$ spinor and $\Nf$ vector 
representations.   We construct a dual to this model based upon an 
$\nobreak{SU(\Nf+2\NQ-7)} \times Sp(2\NQ-2)$ gauge group without utilizing 
deconfinement methods.  This product theory generalizes all previously known 
Pouliot-type duals to $SO(\Nc)$ models with spinor and vector matter.  It also 
yields large numbers of new dual pairs along various flat directions.  The 
dual description of the $SO(10)$ theory satisfies multiple consistency checks 
including an intricate renormalization group flow analysis which links it 
with Seiberg's duality transformations.  We discuss its implications for 
building grand unified theories that contain all Standard Model fields as 
composite degrees of freedom.

\Date{5/97}

\newsec{Introduction}

	During the past two years, significant theoretical interest
and effort has been directed towards finding dual descriptions of
strongly interacting $\CN=1$ supersymmetric gauge theories.  This
enterprise was launched by Seiberg's construction of a dual to SUSY
QCD \SeibergI.  Seiberg's discovery provided valuable insight into
such general nonperturbative aspects of quantum field
theory as confinement, massless solitons, phase transitions and conformal 
fixed points.   A number of duality transformations uncovered since Seiberg's 
pioneering work have shed light upon other interesting phenomena including
strong interaction cloaking of chirality \Pouliot\ and dynamical
supersymmetry breaking \refs{\PouliotStrasslerII{--}\Chou}.
Unfortunately, no systematic field theory method for mapping long distance
universality classes of microscopic supersymmetric gauge theories has
so far been developed, and finding duals to models with more than just 
fundamental matter contents remains difficult.  
\foot{Recent D-brane work of several investigators has led to a deeper 
understanding of $\CN=1$ duality \refs{\Bershadsky{--}\AharonyHanany}.
In this article, we follow a field theory approach to the subject.}
A few specialized strategies which simplify the search have been devised by 
various groups.  But given that the number of such theoretical tools is still 
quite limited, it is clearly worthwhile to discover and study more novel 
examples of duality.

	In this article, we present an entire class of new dual pairs
that exhibit several interesting patterns.  One member of each pair
consists of an $SO(10)$ gauge theory with matter in $\NQ$ spinor
representations and $\Nf$ vector representations.  Its dual
counterpart is a chiral model with semisimple gauge group
$SU(\Nf+2\NQ-7) \times Sp(2\NQ-2)$.
\foot{We take the fundamental irrep of $Sp(2N)$ to be $2N$ dimensional.}
Duals involving product groups have appeared before in the literature
\refs{\IntriligatorThomas{--}\SSBMIT,\IntriligatorSeiberg{--}\Sakai}.
Many of these dualities can be derived by straightforward application of
Seiberg's results \SeibergI\ to individual group factors.  Others
\refs{\BrodieHanany,\ILS,\Brodie} represent nontrivial 
generalizations of the duality transformations of Seiberg and of Kutasov 
\Kutasov, but still connect two theories of the same Cartan class.  For 
example, in such models one finds that an $SU(N)\times SO(N')$ theory with
appropriate matter is dual to a similar $SU(\tilde N)\times 
SO(\tilde N')$ theory.  The product duals which we analyze in this paper
qualitatively differ from these earlier examples.  Moreover, they
generalize all previously known Pouliot-type duals to various
$SO(\Nc)$ gauge theories with particular spinor and vector matter
contents \refs{\Pouliot,\PouliotStrasslerII,\PouliotStrasslerI{--}\ChoI}.  
As we shall see, this new double array of dual pairs provides several novel 
insights into $\CN=1$ duality.

	Our article is organized as follows.  We first focus upon the
$SO(10)$ theory with two spinors and discuss its confining phase in
section~2.  We then construct the $SU(\Nf-3) \times Sp(2)$ dual to the
$\NQ=2$ theory in section~3 and verify that it satisfies multiple
consistency checks including anomaly matching, composite operator
mapping, and duality preservation along flat directions.  In
section~4 which contains our main results, we investigate the general
product duals to $SO(10)$ models with arbitrary numbers of spinor and
vector matter fields.  We use these duals in section~5 to build grand
unified theories that contain all Standard Model bosons and fermions
as composites.  Finally, we summarize our findings in section~6 and present 
details on exotic operator maps, two corollary duality transformations 
and an intricate renormalization group flow analysis in four separate 
appendices.

\newsec{The $\pmb{\NQ=2}$ $\pmb{SO(10)}$ model}

	We begin our study by considering a supersymmetric gauge theory with 
symmetry group 
\eqn\symgroup{G = SO(10)_\local \times \bigl[ SU(\Nf) \times SU(2) 
\times U(1)_\Y \times U(1)_\R \bigr]_\glob,} 
chiral superfields 
\eqn\matter{\eqalign{
V^i_\mu & \sim \bigl( 10; \fund ,1 ; -4, \Rtwo \bigr) \cr
Q^\A_\I  & \sim \bigl( 16; 1,2 ; \Nf, \Rtwo \bigr) \cr}}
and zero tree level superpotential.  The hypercharge and R-charge assignments 
for the vector and spinor matter fields are chosen so that the model is free 
of global anomalies.  It is also asymptotically free so long as its Wilsonian 
beta function coefficient
\foot{We adopt the $SO(10)$ index values $K(10) = 2$, $K(16)=4$, and 
$K(45) = 16$.}
\eqn\betaWilson{b_0 = \half \bigl[ 3 K({\rm Adj}) -
\sum_{\rm{\buildrel matter \over {\scriptscriptstyle reps \>\rho}}} K(\rho)
\bigr] = 20 - \Nf}
is positive.  The beta function that governs the running of the
physical gauge coupling is then negative at weak coupling \Shifman.  The
theory's infrared dynamics are consequently nontrivial provided it
contains $\Nf < 20$ vectors.  

	Generic expectation values for the matter fields break the $SO(10)$ 
gauge group according to the pattern \Elashvili\
\eqn\pattern{SO(10) \> {\buildrel 2(16) \over \longrightarrow} \>
               G_2 \> {\buildrel 10 \over \longrightarrow} \> 
               SU(3) \> {\buildrel 10 \over \longrightarrow} \>
               SU(2) \> {\buildrel 10 \over \longrightarrow} \>
               1.}
This expression illustrates the hierarchy of gauge symmetries realized
at progressively longer distance scales, assuming that the spinor vevs' 
magnitudes are larger than the first vector's, which in turn is larger than 
the second vector's, and so forth.  Using this symmetry breaking information, 
we can straightforwardly count the gauge invariant operators that are needed 
to act as moduli space coordinates for small numbers of vector flavors 
\refs{\Pouliot,\ChoII}.  Such gauge singlets enter
into the effective low energy description of the microscopic $SO(10)$
theory.  In Table~1, we display the number of partonic degrees of
freedom and the generic unbroken color subgroup $H_\local$ as a
function of $\Nf$.  The dimension of the coset space $G_\local /
H_\local$ coincides with the number of matter fields eaten by the
superHiggs mechanism.  The number of remaining uneaten partons listed
in the last column of the table equals the number of independent
hadrons which label flat directions in the effective theory.

\midinsert
\parasize=1in

\begintable
$\quad \Nf \quad $ \| Parton DOF \| Unbroken Subgroup \| Eaten DOF
\| Hadrons \crthick
0 \| 32 \| $G_2$   \| $45-14=31$ \| 1 \nr
1 \| 42 \| $SU(3)$ \| $45-8=37$ \| 5 \nr
2 \| 52 \| $SU(2)$ \| $45-3=42$ \| 10 \nr
3 \| 62 \| 1 \| 45 \| 17 \nr
4 \| 72 \| 1 \| 45 \| 27 \nr 
5 \| 82 \| 1 \| 45 \| 37 \endtable
\bigskip
\centerline{Table 1: Number of independent hadrons in the $SO(10)$
theory with two spinors}
\bigskip
\endinsert

	In order to explicitly construct the hadron fields, it is useful to 
recall the tensor decomposition of the $SO(10)$ spinor product
\eqn\spinorprod{16 \times 16 = 10_\S + 120_\A + 126_\S 
			     = [1]_\S + [3]_\A + \widetilde{[5]}_\S.}
Here $[n]$ denotes an antisymmetric rank-$n$ tensor irrep, the ``S'' and ``A'' 
subscripts indicate symmetry and antisymmetry under spinor exchange and the 
tilde over the last term implies that the rank-5 irrep is complex self-dual 
\Harvey.  We form symmetric and
antisymmetric combinations of the two spinors using the Pauli matrices
$\s_\xrm \s_2 \> (\xrm=1,2,3)$ and $\s_2$ as Clebsch-Gordan
coefficients.  We next contract vector superfields into the bispinor
pairs utilizing $SO(10)$ Gamma matrices $\Gamma_\mu$ and charge
conjugation matrix $C$.
\foot{We implicitly regard the 16-dimensional $SO(10)$ spinor as the 
projection $Q=P_- Q_{32}$ where $Q_{32}$ denotes the 32-dimensional spinor 
of $SO(11)$ and $P_- = \half (1-\Gamma_{11})$.  The $32 \times 32$ Gamma 
matrices of $SO(10)$ along with its charge conjugation matrix $C$ come 
from the Clifford algebra of $SO(11)$.}
We thus produce the gauge invariant composites 
\eqn\eleccomposites{\eqalign{
K &= Q^\T_\I (\s_\xrm \s_2)_{\I\J} \Gamma^\mu C Q_\J 
  Q^\T_\K (\s_\xrm \s_2)_{\K\L} \Gamma_\mu C Q_\L 
  \sim \bigl(1; 1, 1; 4\Nf, 4 \Rtwo \bigr) \cr
M^{(ij)} &= (V^\T)^{i \mu} V^j_\mu 
	\sim \bigl(1; \sym, 1; -8, 2 \Rtwo \bigr) \cr
N^i_\xrm &= Q^\T_\I (\s_\xrm \s_2)_{\I\J} \Vslash^i C Q_\J 
	\sim \bigl(1; \fund,3; 2 \Nf - 4, 3 \Rtwo \bigr) \cr
P^{[ijk]} &= {1 \over 3!} Q^\T_\I (\s_2)_{\I\J} \Vslash^{[i} \Vslash^j 
  \Vslash^{k]} C Q_\J
  \sim \bigl(1; \antithree,1; 2\Nf - 12, 5 \Rtwo \bigr) \cr
R^{[ijkl]} &= {1 \over 4!} Q^\T_\I (\s_\xrm \s_2)_{\I\J} \Gamma^\mu C Q_\J
	Q^\T_\K (\s_\xrm \s_2)_{\K\L} \Gamma_\mu \Vslash^{[i} 
  \Vslash^j \Vslash^k \Vslash^{l]} C Q_\L
\sim \bigl(1; \antifour, 1; 4\Nf - 16, 8 \Rtwo \bigr)\cr 
T^{[ijklm]}_\xrm &= {1 \over 5!} Q^\T_\I (\s_\xrm \s_2)_{\I\J} 
  \Vslash^{[i} \Vslash^j \Vslash^k \Vslash^l \Vslash^{m]} C Q_\J
	\sim \bigl(1; \antifive,3; 2\Nf - 20, 7 \Rtwo \bigr) \cr}}
where Greek, small Latin and large Latin letters respectively denote $SO(10)$ 
color, $SU(\Nf)$ vector and $SU(2)$ spinor indices. 

	It is instructive to compare the number of these composite operators
with the number of independent flat directions as a function of $\Nf$.  We 
perform this comparison in Table~2.  Looking at the 
table's entries, we see that $K$, $M$, $N$ and $P$ account for all massless 
fields in the $SO(10)$ model with three or fewer vector flavors.  On the other 
hand, the hadron count exceeds the needed number of composites by one when 
$\Nf=4$.  A single constraint must therefore exist among the hadron fields in 
this case.  The precise quantum constraint relation is fixed by symmetry and 
the classical limit \SeibergII.  It appears in superpotential form as 
\eqn\Wconstraint{\eqalign{
W_{\Nf=4} &= X \bigl[ 4 R^2 + 12 K P_i M^{ij} P_j -36 N^i_\xrm N^j_\xrm P_i P_j 
  - 12 i \e_{i_1 i_2 i_3 i_4} \exyz N^{i_1}_\xrm N^{i_2}_\yrm N^{i_3}_\zrm 
  M^{i_4 j} P_j \cr
& \qquad + \e_{i_1 i_2 i_3 i_4} \e_{j_1 j_2 j_3 j_4}
\bigl( 2 K M^{i_1 j_1} M^{i_2 j_2} M^{i_3 j_3} N^{i_4}_\xrm  N^{j_4}_\xrm 
- 9 N^{i_1}_\xrm N^{j_1}_\xrm N^{i_2}_\yrm N^{j_2}_\yrm M^{i_3 j_3} M^{i_4 j_4}
\bigr) \cr
& \qquad -4 K^2 \det M - \Lambda_4^{16} \bigr]  \cr}}
where the superfield $X$ represents a Lagrange multiplier. As a check on this 
result, one can Higgs the $SO(10)$ gauge group down to $SO(6) \simeq SU(4)$ by 
giving vevs $\langle V^i_\mu \rangle = v \d^i_\mu$ to the four vectors.  The 
two 16-dimensional spinor fields then split apart into four $4+\bar{4}$ pairs.  
After the $SO(10)$ hadrons are decomposed in terms of $SU(4)$ mesons and 
baryons, one finds that the quantum constraint in \Wconstraint\ properly 
reduces to the $\Nf=\Nc=4$ SUSY QCD relation $\det M_{4 \times 4} - B \bar{B} 
= \Lambda^8$ \SeibergII.

\topinsert
\parasize=1in

\begintable
$\quad \Nf \quad $ \| Hadrons \| $ \sp K \sp$ \| $\sp M \sp$ \| $\sp N \sp$
\| $\sp P  \sp$ \| $\sp R  \sp$ \| $\sp T \sp$ \| constraints \crthick
0 \| 1  \| 1 \|    \|   \|     \|   \|   \|     \cr
1 \| 5  \| 1 \| 1  \| 3 \|     \|   \|   \|     \cr
2 \| 10 \| 1 \| 3  \| 6 \|     \|   \|   \|     \cr
3 \| 17 \| 1 \| 6  \| 9 \|  1  \|   \|   \|     \cr
4 \| 27 \| 1 \| 10 \| 12 \| 4  \| 1 \|   \| -1  \cr
5 \| 37 \| 1 \| 15 \| 15 \| 10 \| 5 \| 3 \| -12 \endtable
\bigskip
\centerline{Table 2: Hadron degree of freedom count in the $SO(10)$ theory 
  with two spinors}
\bigskip
\endinsert

	The hadron fields in the $\Nf=5$ theory are restricted by 12 
independent relations.  These constraints are encoded 
within the equations of motion that follow from the $\Nf=5$ superpotential
\eqn\Wconfining{\eqalign{
& W_{\Nf=5} = {1 \over {\Lambda_5^{15}}} \Bigl[ 12 K T_\xrm T_\xrm
  + 36 i \exyz N^i_\xrm N^j_\yrm P_{ij} T_\zrm
  -24 N^i_\xrm R_i T_\xrm + 3 \e^{ijklm} P_{ij} P_{kl} R_m
  + 4 R_i M^{ij} R_j \cr
& \> + 6 K M^{ij} M^{kl} P_{ik} P_{jl}
  - 36 M^{ij} N^k_\xrm N^l_\xrm P_{ik} P_{jl}
  - 6 i \e_{i_1 i_2 i_3 i_4 i_5} \exyz N^{i_1}_\xrm N^{i_2}_\yrm
  N^{i_3}_\zrm M^{i_4 j} M^{i_5 k} P_{jk} \cr
& \> + \e_{i_1 i_2 i_3 i_4 i_5} \e_{j_1 j_2 j_3 j_4 j_5}
 \bigl(\half K M^{i_1 j_1} M^{i_2 j_2} M^{i_3 j_3} M^{i_4 j_4} N^{i_5}_\xrm
  N^{j_5}_\xrm - 3 N^{i_1}_\xrm N^{j_1}_\xrm N^{i_2}_\yrm N^{j_2}_\yrm
  M^{i_3 j_3} M^{i_4 j_4} M^{i_5 j_5} \bigr) \cr
& \> -4 K^2 \det M \bigr]. \cr}}
After varying this complicated expression, we observe that the origin 
$K=M=N=P=R=T=0$ satisfies all equations of motion and lies on the quantum 
moduli space.  Since the full global symmetry group remains unbroken at this 
point, all `t~Hooft anomalies calculated within the microscopic and effective 
theories should agree.  We find that the parton and hadron level 
$SU(\Nf)^3$, $SU(\Nf)^2 U(1)_\Y$, $SU(\Nf)^2 U(1)_\R$, $SU(2)^2 U(1)_\Y$, 
$SU(2)^2 U(1)_\R$, $U(1)_\Y$, $U(1)_\Y^3$, $U(1)_\R$, $U(1)_\R^3$, 
$U(1)_\Y^2 U(1)_\R$ and $U(1)_\R^2 U(1)_\Y$ anomalies do indeed match when 
$\Nf=5$ \CSS.  This nontrivial anomaly agreement strongly suggests that the 
effective theory contains only the composite fields in \eleccomposites\ and no 
additional colored or colorless massless degrees of freedom.  

	The $\NQ=2$, $\Nf=5$ $SO(10)$ model is clearly analogous to
$\Nf=\Nc+1$ SUSY QCD \SeibergII\ which is commonly, though
imprecisely, referred to as confining.  The $SO(10)$ theory with
$\NQ=2$, $\Nf=6$ is similarly reminiscent of SUSY QCD with
$\Nf=\Nc+2$ flavors, inasmuch as microscopic and macroscopic global
anomalies fail to match and any effective superpotential would have to 
involve a branch cut.  The anomaly mismatch cannot be remedied by introducing 
additional color-singlet fields without disrupting the $\Nf=5$ results.  
Moreover, the $N_f=6$ model flows along a spinor flat direction to $G_2$ 
theory with six fundamentals which is known to exist within a nonabelian 
Coulomb phase \Pouliot.  We conclude that the $\Nf=6$ $SO(10)$ theory cannot 
be represented in terms of gauge singlet operators in the far infrared.  
Instead, we expect that it resides within a new phase which possesses a dual 
description.   As we shall see in the next section, this expectation is 
correct.

\newsec{The $\pmb{SU(\Nf-3) \times Sp(2)}$ dual}

        Although no systematic field theory method currently exists for 
constructing nontrivial duals, some patterns have been found in special cases.  
For instance, partners to several models which possess tree level 
superpotentials \refs{\ILS,\Brodie,\KutasovSchwimmer{--}\KSS} have been 
uncovered following Kutasov's first example of SUSY QCD with adjoint matter 
\Kutasov. Another class of dual pairs which exhibits certain trends stems 
from Pouliot's $G_2$ model with $\Nf$ fields in the fundamental irrep 
\Pouliot.  The dual to this theory is based upon an $SU(\Nf-3)$ gauge group, 
and its matter content includes a symmetric tensor.  Counterparts to $SO(7)$, 
$SO(8)$, $SO(9)$ and $SO(10)$ generalizations of Pouliot's $G_2$ model are 
qualitatively similar
\refs{\Pouliot,\PouliotStrasslerII,\PouliotStrasslerI{--}\ChoI}.  Unlike 
the Kutasov-type dual pairs, the Pouliot-type examples can be 
arranged to have zero tree level superpotential on one side.  Of course after 
having found such transformations, one can always choose to turn on 
some classical superpotential and study resulting dual pair deformations. 

	Our $SO(10)$ theory with two spinors (which we will refer to
as the ``electric'' theory) reduces to Pouliot's $G_2$ model along a flat 
direction where both spinors acquire vevs.  We consequently begin our search 
for a dual to the $\NQ=2$ model (which we will refer to as the ``magnetic'' 
theory) by looking for extensions of the $SU(\Nf-3)$ counterpart to the
$G_2$ theory.  Since Higgsing the electric theory often induces mass 
decoupling on the magnetic side but leaves the color group unaltered, we 
hypothesize that $\Gdual$ contains a local $SU(\Nf-3)$ factor.  We also 
presume that the $SO(10)$ model's full $SU(\Nf) \times SU(2) \times U(1)_\Y 
\times U(1)_\R$ global symmetry is realized at short as well as long distance 
scales in the dual.  Our initial guess for the magnetic symmetry group is thus
\eqn\maggroupguess{\Gdual = \bigl[SU(\Nf-3) \bigr]_\local 
\times \bigl[ SU(\Nf) \times SU(2) \times U(1)_\Y \times U(1)_\R 
\bigr]_\glob.} 

	Following the examples of all previously constructed Pouliot-type 
duals, we introduce quark matter fields which transform under $\Gdual$ as 
\eqn\dualquarks{q^\a_i \sim \bigl( \fund ; \antifund,1; 
2{\Nf-6 \over \Nf-3} , {5 (\Nf-4)  \over (\Nf-3)(\Nf+4)} \bigr).}
The hypercharge and R-charge assignments have been chosen so that the 
magnetic baryon $p=q^{\Nf-3}$ naturally maps onto the electric baryon $P$ in 
\eleccomposites.  If we also incorporate the composite $M^{(ij)}$ and 
$N^i_\xrm$ electric mesons into the magnetic theory as elementary fields 
$m^{(ij)}$ and $n^i_\xrm$, we find that the $SU(\Nf)^3$, $SU(\Nf)^2 U(1)_\Y$ 
and $SU(\Nf)^2 U(1)_\R$ global `t~Hooft anomalies match between the $SO(10)$ 
model and its dual.  This nontrivial anomaly agreement suggests that 
we have properly identified all magnetic matter fields which transform 
nontrivially under the $SU(\Nf)$ global symmetry group.

	We next introduce into the magnetic theory a symmetric tensor field 	
\eqn\symmetrictensor{s_{\a\b} \sim \bigl( \symbar; 1,1; {4\Nf  \over \Nf-3}, 
2 {3 \Nf-4  \over (\Nf-3)(\Nf+4)} \bigr),}
dual antiquarks
\eqn\dualantiquarks{\qbar^\xrm_\a \sim \bigl( \antifund; 1,3; 
-2\Nf{\Nf-4 \over \Nf-3}, - {\Nf^2-18 \Nf + 40 \over (\Nf-3)(\Nf+4)} \bigr)}
and superpotential interaction terms
\eqn\Wmagneticshort{
\Wmag = {1 \over \mu_1^2} m^{(ij)} q^\a_i s_{\a\b} q^\b_j + 
  {1 \over \mu_2^2} n^i_\xrm  q^\a_i \qbar^\xrm_\a.}
All of these items are common ingredients in Pouliot-type duals.  The abelian 
quantum numbers for $s$ and $\qbar$ are fixed by requiring invariance of both 
terms in $\Wmag$ under $\Gdual$.  With these dual matter fields now 
in hand, we find that the electric baryons 
\foot{The two anticommuting fermionic field strength tensors in $B_2$, $C_2$, 
$b_0$ and $c_0$ must be contracted together into antisymmetric spin-0 
combinations in order for these electric and magnetic baryons not to vanish.}
\eqn\elecbaryons{\eqalign{
B_0^{[i_1 \cdots i_{10}]} &= \e^{\mu_1 \cdots \mu_{10}} V^{i_1}_{\mu_1} \cdots 
  V^{i_{10}}_{\mu_{10}} \cr
B_1^{a[i_1 \cdots i_8]} &= \e^{\mu_1 \cdots \mu_{10}} V^{i_1}_{\mu_1} \cdots
  V^{i_8}_{\mu_8} W^a_{\mu_9 \mu_{10}} \cr
B_2^{[i_1 \cdots i_6]} &= \e^{\mu_1 \cdots \mu_{10}} \e_{ab} 
V^{i_1}_{\mu_1} \cdots V^{i_6}_{\mu_6} W^a_{\mu_7 \mu_8} W^b_{\mu_9 \mu_{10}} 
\cr}}
can be mapped onto the magnetic operators
\eqn\magantibaryons{\eqalign{
b_0^{[i_1 \cdots i_{10}]} &= \e^{i_1 \cdots i_\Nf} \e^{\a_1 \cdots \a_{\Nf-3}} 
\exyz \e_{ab} 
\cr
&\quad\quad
\sp (sq)_{\a_1 i_{11}} \cdots (sq)_{\a_{\Nf-10} i_\Nf} 
(s \wtilde^a)_{\a_{\Nf-9} \a_{\Nf-8}} 
(s \wtilde^b)_{\a_{\Nf-7} \a_{\Nf-6}} 
 \qbar^\xrm_{\a_{\Nf-5}} \qbar^\yrm_{\a_{\Nf-4}} 
\qbar^\zrm_{\a_{\Nf-3}} \cr
b_1^{a[i_1 \cdots i_8]} &= \e^{i_1 \cdots i_\Nf} \e^{\a_1 \cdots \a_{\Nf-3}} 
\exyz \sp 
\cr
&\quad\quad (sq)_{\a_1 i_9} \cdots (sq)_{\a_{\Nf-8} i_\Nf} 
(s \wtilde^a)_{\a_{\Nf-7} \a_{\Nf-6}} 
\qbar^\xrm_{\a_{\Nf-5}} \qbar^\yrm_{\a_{\Nf-4}} 
\qbar^\zrm_{\a_{\Nf-3}} \cr
b_2^{a[i_1 \cdots i_6]} &= \e^{i_1 \cdots i_\Nf} \e^{\a_1 \cdots \a_{\Nf-3}} 
\exyz \sp (sq)_{\a_1 i_7} \cdots (sq)_{\a_{\Nf-6} i_\Nf} 
\qbar^\xrm_{\a_{\Nf-5}} \qbar^\yrm_{\a_{\Nf-4}} 
\qbar^\zrm_{\a_{\Nf-3}}.  \cr}}
The more exotic combinations of $SO(10)$ vectors, spinors and gluons
\eqn\elecexotics{\eqalign{
{C_0}_\xrm^{[i_1 \cdots i_9]} &= \e^{\mu_1 \cdots \mu_{10}} 
V^{i_1}_{\mu_1} \cdots V^{i_9}_{\mu_9} Q^\T_\I (\s_\xrm \s_2)_{\I\J} 
\Gamma_{\mu_{10}}   C Q_\J \cr
{C_1}_\xrm^{a [i_1 \cdots i_7]} &= \e^{\mu_1 \cdots \mu_{10}} V^{i_1}_{\mu_1} 
  \cdots V^{i_7}_{\mu_7} W^a_{\mu_8 \mu_9} Q^\T_\I (\s_\xrm \s_2)_{\I\J} 
  \Gamma_{\mu_{10}} C Q_\J \cr
{C_2}_\xrm^{[i_1 \cdots i_5]} &= \e^{\mu_1 \cdots \mu_{10}} \e_{ab} 
V^{i_1}_{\mu_1} \cdots V^{i_5}_{\mu_5} W^a_{\mu_6 \mu_7} W^b_{\mu_8 \mu_9} 
  Q^\T_\I (\s_\xrm \s_2)_{\I\J} \Gamma_{\mu_{10}} C Q_\J \cr}}
may similarly be identified with the dual composites
\eqn\magexotics{\eqalign{
{c_0}_\xrm^{[i_1 \cdots i_9]} &= \e^{i_1 \cdots i_\Nf} 
\e^{\a_1 \cdots \a_{\Nf-3}} \exyz \e_{ab} 
\cr
&\quad\quad  
\sp (sq)_{\a_1 i_{10}} \cdots (sq)_{\a_{\Nf-9} i_\Nf} 
(s \wtilde^a)_{\a_{\Nf-8} \a_{\Nf-7}} 
(s \wtilde^b)_{\a_{\Nf-6} \a_{\Nf-5}} 
 \qbar^\yrm_{\a_{\Nf-4}} \qbar^\zrm_{\a_{\Nf-3}} \cr
{c_1}_\xrm^{a [i_1 \cdots i_7]} &= \e^{i_1 \cdots i_\Nf} 
\e^{\a_1 \cdots \a_{\Nf-3}} \exyz
  \sp (sq)_{\a_1 i_8} \cdots (sq)_{\a_{\Nf-7} i_\Nf} 
(s \wtilde^a)_{\a_{\Nf-6} \a_{\Nf-5}} 
\qbar^\yrm_{\a_{\Nf-4}} \qbar^\zrm_{\a_{\Nf-3}} \cr
{c_2}_\xrm^{[i_1 \cdots i_5]} &= \e^{i_1 \cdots i_\Nf} 
\e^{\a_1 \cdots \a_{\Nf-3}} \exyz
  \sp (sq)_{\a_1 i_6} \cdots (sq)_{\a_{\Nf-5} i_\Nf} 
\qbar^\yrm_{\a_{\Nf-4}} \qbar^\zrm_{\a_{\Nf-3}} .\cr}}
The overall consistency of these gauge invariant operator identifications is 
encouraging.

	Our construction of the dual to the $SO(10)$ theory with two spinors 
has so far closely mimicked that for the dual to $SO(10)$ with one spinor 
\PouliotStrasslerII.  But as other investigators have recently observed 
\DistlerKarch, continuing in this direction ultimately leads to a dead-end.  
We cannot find a magnetic theory based upon the symmetry group in 
\maggroupguess\ for which all local anomalies cancel, all global anomalies 
match and all composite operators map.  We must therefore relax some 
assumption in order to make further progress.  After exploring several 
possibilities, we are forced to conclude that the magnetic gauge group 
is not simple.  We consequently expand our search by looking for a dual with 
a product color group.  

	The simplest generalization which retains the previous desirable 
features while overcoming the above-mentioned difficulties has symmetry group
\foot{We remind the reader that in our conventions $Sp(2) \simeq SU(2)$.
We use dotted Greek letters to denote $Sp(2)$ color indices.}
\eqn\maggroup{\Gdual = \bigl[SU(\Nf-3) \times Sp(2) \bigr]_\local 
\times \bigl[ SU(\Nf) \times SU(2) \times U(1)_\Y \times U(1)_\R 
\bigr]_\glob,} 
superfield matter content
\eqn\magmatter{\eqalign{
q^\a_i &\sim \bigl( \fund ,1 ; \antifund,1; 2{\Nf-6 \over \Nf-3} , 
  {5 (\Nf-4) \over (\Nf-3)(\Nf+4)} \bigr) \cr
\qp^{\a\adot}_\I &\sim \bigl( \fund,2; 1,2; -{2 \Nf \over \Nf-3}, 
  {(\Nf+2)(\Nf-4) \over (\Nf-3)(\Nf+4)} \bigr) \cr
\qbar^\xrm_\a &\sim \bigl( \antifund,1; 1,3; -2\Nf{\Nf-4 \over \Nf-3}, 
  - {\Nf^2-18 \Nf + 40 \over (\Nf-3)(\Nf+4)} \bigr) \cr
s_{\a\b} &\sim \bigl( \symbar,1; 1,1; {4\Nf  \over \Nf-3}, 
  2 {3 \Nf-4  \over (\Nf-3)(\Nf+4)} \bigr) \cr
t^{\adot}_\I &\sim \bigl( 1,2; 1,2; 2\Nf, 2 \Rtwo \bigr) \cr
m^{(ij)} &\sim \bigl(1,1; \sym,1; -8, 2 \Rtwo \bigr) \cr
n^i_\xrm  &\sim \bigl(1,1; \fund,3; 2\Nf-4, 3 \Rtwo \bigr) \cr}}
and tree level superpotential
\eqn\Wmagnetic{
\Wmag = {1 \over \mu_1^2} m^{(ij)} q^\a_i s_{\a\b} q^\b_j 
 + {1 \over \mu_2^2} n^i_\xrm  q^\a_i \qbar^\xrm_\a 
 + \lambda_1 \e_{\adot \bdot} \e^{\I\J} {\qp}^{\a\adot}_\I s_{\a\b} 
    {\qp}^{\b\bdot}_\J
 + \lambda_2 \e_{\adot \bdot} {\qp}^{\a\adot}_\I (\s_\xrm \s_2)^{\I\J} 
  \qbar^\xrm_\a t^{\bdot}_\J.}
Several points about this product dual should be noted.  Firstly, the 
$SU(\Nf-3)^3$, $SU(\Nf-3)^2 U(1)_\Y$, $SU(\Nf-3)^2 U(1)_\R$, 
$Sp(2)^2 U(1)_\Y$, and $Sp(2)^2 U(1)_\R$ anomalies vanish, and an even number 
of doublets transform under the color $Sp(2)$ \Witten.  The magnetic gauge 
group and global abelian hypercharge and R-charge symmetries are consequently 
free of all perturbative and nonperturbative anomalies like their electric 
counterparts.  All anomalies involving only global symmetries also match 
between the electric and magnetic theories for $\nobreak{\Nf \ge 5}$.  
Secondly, we treat the colored partons in this theory as canonically 
normalized, but we set the engineering mass dimensions of the colorless 
$m^{(ij)}$ and $n^i_\xrm$ fields equal to those of their electric 
counterparts.  In order for the magnetic superpotential to have dimension 
three, the first two nonrenormalizable terms in \Wmagnetic\ must be multiplied 
by dimensionful prefactors $\mu_1^{-2}$ and $\mu_2^{-2}$.  On the other hand, 
the $\lambda_1$ and $\lambda_2$ coefficients in the third and fourth terms 
of $\Wmag$ are dimensionless.  For simplicity, we set all these prefactors 
equal to unity from here on.  Finally, it is instructive to count the number 
of constraints on abelian charge assignments which determines how many 
nonanomalous $U(1)$ factors appear within the magnetic global symmetry group.  
We start with the 7 fermionic components of the $q$, $\qp$, $\qbar$, $s$, $t$, 
$m$, and $n$ matter superfields along with the $SU(\Nf-3)$ gluino $\tilde{g}$ 
and $Sp(2)$ wino $\tilde{w}$.  Requiring the $SU(\Nf-3)^2 U(1)$ and 
$Sp(2)^2 U(1)$ anomalies to vanish imposes two conditions on these fermions' 
charges.  We next recall that fermion-sfermion-gaugino interactions in the
Kahler potential tie together the global quantum numbers for $\tilde{g}$ and 
$\tilde{w}$.  The terms in the magnetic superpotential \Wmagnetic\ impose 4 
more conditions on abelian charge assignments.  We thus find that the magnetic 
dual possesses $9-2-1-4=2$ independent global $U(1)$ symmetries which 
agrees with the electric theory.

	The dual pair's phase structure represents an important 
dynamical issue. The Wilsonian beta function coefficients for the two 
gauge groups in $\Gdual$ are given by $\tilde{b_0}^{SU(\Nf-3)} = 
2 \Nf -12$ and $\tilde{b_0}^{Sp(2)} = 8 - \Nf$.  The $SU(\Nf-3)$ factor is 
asymptotically free for $\Nf \ge 7$, while the $Sp(2)$ factor is 
asymptotically free for $\Nf \le 7$. Since there is no value of $\Nf$ for 
which both are free in the infrared,  the $SO(10)$ theory does not possess a 
free magnetic phase. Instead, it exists at the origin of moduli space in a 
nonabelian Coulomb phase for $6 \le \Nf \le 19$ vector flavors.  For 
$N_f \ge 20$, the magnetic theory flows to the weakly coupled $SO(10)$ theory 
at long distance scales.  The absence of a free magnetic phase is a common 
feature in all similar dual pairs 
\refs{\Pouliot,\PouliotStrasslerII,\PouliotStrasslerI{--}\ChoI}.  

	We next construct maps between gauge invariant operators in the 
$SO(10)$ and $SU(\Nf-3) \times Sp(2)$ theories.  We have already matched 
several operators in \elecbaryons--\magexotics, but there are many others to 
consider.  We first identify the composites
\eqna\magbaryons
$$ \eqalignno{ 
k &= \e_{\adot \bdot} \e^{\I\J} t^{\adot}_\I t^{\bdot}_\J & \magbaryons a \cr
p^{[i_1 i_2 i_3]} &= \e^{i_1 \cdots i_\Nf} \e_{\a_1 \cdots \a_{\Nf-3}} 
 \sp q^{\a_1}_{i_4} \cdots q^{\a_{\Nf-3}}_{i_\Nf} & \magbaryons b \cr
r^{[i_1 i_2 i_3 i_4]}  &= \e^{i_1 \cdots i_\Nf} \e_{\a_1 \cdots \a_{\Nf-3}} 
  \e_{\adot \bdot} \e_{\I\J}
  \sp q^{\a_1}_{i_5} \cdots q^{\a_{\Nf-4}}_{i_\Nf} \qp^{\a_{\Nf-3} \adot}_\I 
  t^{\bdot}_\J & \magbaryons c \cr
t^{[i_1 i_2 i_3 i_4 i_5]}_\xrm &= \e^{i_1 \cdots i_\Nf} 
\e_{\a_1 \cdots \a_{\Nf-3}} \e_{\adot \bdot} (\s_\xrm \s_2)_{\I\J} 
 \sp q^{\a_1}_{i_6} \cdots q^{\a_{\Nf-5}}_{i_\Nf} \qp^{\a_{\Nf-4} \adot}_\I 
\qp^{\a_{\Nf-3} \bdot}_\J & \magbaryons d \cr} $$
as partners to $K$, $P$, $R$ and $T$ in \eleccomposites.  These magnetic 
hadrons share exactly the same quantum numbers as their electric theory 
counterparts.  In particular, their transformation rules under the global 
$SU(2)$ which rotates the two spinors on the electric side are fixed once we 
form gauge invariant combinations of the dual matter fields.  Other chiral 
operators besides those which act as moduli space coordinates in the confining 
phase can also be mapped.  For example, the exotic $SO(10)$ invariants
\eqn\moreelecexotics{\eqalign{
U &= \e_{ab} Q_1 \Gamma^{[\mu_1\mu_2\mu_3]} Q_2 
Q_1 \Gamma^{[\mu_1\nu_2\nu_3]} Q_2 W^a_{\mu_2\nu_2} W^b_{\mu_3\nu_3} \cr
V &= \e_{ab} Q_1 \Gamma^{[\mu_1\mu_2\mu_3]} Q_2 
Q_1 \Gamma^{[\mu_1\mu_2\nu_3]} Q_2 W^a_{\mu_3\sigma} W^b_{\nu_3\sigma} \cr}}
are identified with linear combinations of the $SU(\Nf-3) \times Sp(2)$ 
singlets
\eqn\moremagexotics{\eqalign{
u & = \det s \cr
v & = \e_{\adot\bdot} \e_{\cdot \ddot} \e^{\I\J} \e^{\K\L} 
\qp^{\a\adot}_\I t^{\bdot}_\J s_{\a\b} \qp^{\b\cdot}_\K t^{\ddot}_\L. \cr}}
We should point out that $\det s$ does not appear anywhere within the magnetic 
superpotential \Wmagnetic.  This feature of our product dual represents an 
interesting departure from previously studied Pouliot-type models.  Since 
$u=\det s$ is not rendered redundant by equations of motion, it must match onto 
some linear combination of the primary $U$ and $V$ electric composites.  
Further details on mapping these $Q^4 W^2$ operators are presented in 
Appendix~A.

	Given that $\det s$ does not appear in the $SU(\Nf-3) \times Sp(2)$ 
theory's superpotential but does enter into $\Wmag$ in Pouliot's $SU(\Nf-3)$ 
dual to the $G_2$ model, we should inquire how this term arises when 
we deform the former into the latter.  Recall from \pattern\ that generic 
expectation values for the two 16-dimensional spinors $Q_1$ and $Q_2$ break 
the $SO(10)$ color group down to $G_2$.  This symmetry breaking can 
alternatively be viewed as resulting from a large expectation value for 
the composite operator $K \sim (Q_1 Q_2)^2$: 
\eqn\Kvev{\vev{K} = a^4 \gg \Lambda_{SO(10)}^4.}
The mapping in \magbaryons{a}\ then implies that the dual parton field $t$ 
develops a nonzero vev which can be rotated into the form 
$\vev{t^\adot_\I} = a^2 \e^\adot_\I$.   This condensate for $t$ gives mass 
to the $\qp$ and $\qbar$ matter fields via the last term in \Wmagnetic.  Once 
heavy degrees of freedom are integrated out, the tree level magnetic 
superpotential reduces to 
\eqn\Wmagred{\Wmag \to {m'}^{(ij)} q^\a_i s_{\a\b} q^\b_j - 
  2 q_0^\a s_{\a\b} q_0^\b}
where ${m'}^{(ij)} \equiv m^{(ij)} - {1 \over 2 a^4} n^i_\xrm n^j_\xrm$ and
$q_0^\a$ denotes the only component of $\qp^{\a\adot}_\I$ which does not 
grow massive.  The vev for $t$ also completely breaks the magnetic $Sp(2)$ 
color group.  As a result, instanton effects generate the dynamical 
superpotential \ADSI\ 
\foot{The nonperturbative mechanism underlying (3.18) is the same as that 
which produces the determinant within the $\Nf=\Nc+1$ SUSY QCD superpotential 
$\nobreak{W = (B M \bar{B} - \det M) /\Lambda^{2\Nc-1}}$ \SeibergII.  When the 
magnetic $SU(2)$ gauge group in the $\Nf=\Nc+2$ theory is Higgsed, the 
$\det M$ term is generated by a weak coupling instanton process \SeibergI.}
\eqn\Winstanton{\Wdyn = {\Lambda_{Sp(2)}^{8-\Nf} \over \vev{K}} \det s.}
After combining this quantum contribution with the classical terms in 
\Wmagred, we reproduce the total superpotential in Pouliot's dual to the $G_2$ 
model along with its gauge group and matter content \Pouliot.  This successful 
recovery of an old dual from our new one constitutes an important consistency 
check.

	It is interesting to explore other deformations of the $\NQ=2$ 
$SO(10)$ model which yield novel dual pairs.  We can find magnetic 
descriptions for a large number of electric theories by flowing along various 
vector and spinor flat directions as illustrated in \flowdiag.  For example, we 
sketch the derivation of a dual to an $SO(7)$ model with $\Nf$ spinors and one 
vector.  We first Higgs the $SO(10)$ gauge group down to $SO(7)$ by turning 
on a vev for $N_+^\Nf \equiv N_1^\Nf + i N_2^\Nf$.  We next give mass to all 
singlets not eaten by the superHiggs mechanism.  The resulting deformed 
electric theory becomes 
\eqn\sosevensymgrp{G = SO(7)_\local \times \bigl[ SU(\Nf) \times U(1)_\Y 
\times U(1)_\R \bigr]_\glob} 
with superfield matter content 
\eqn\matter{\eqalign{
V_\mu & \sim \bigl( 7; 1; \Nf,  {\Nf-4 \over \Nf+1} \bigr) \cr
Q^\A_\I  & \sim \bigl( 8; \fund; -1, {\Nf-4 \over \Nf+1} \bigr).  \cr}}
On the magnetic side of the dual pair, we rename all matter fields that 
previously transformed as doublets or triplets under the global $SU(2)$ in 
\maggroup\ which no longer exists in the new $SO(7)$ theory.  Then after 
inserting $\vev{n_+^\Nf}$ into $\Wmag$ and eliminating heavy degrees of 
freedom by solving their equations of motion, we find that the deformed dual 
has symmetry group 
\eqn\sosevendualsymgrp{\Gdual = \bigl[SU(\Nf-3) \times Sp(2) \bigr]_\local 
\times \bigl[ SU(\Nf-1) \times U(1)_\Nf \times U(1)_\Y \times U(1)_\R 
\bigr]_\glob,}
%
%
matter content 
\eqn\magmatter{\eqalign{
q^\a_i &\sim \bigl( \fund ,1 ; \antifund,{2 \over \Nf-3}, {\Nf-2 \over \Nf-3}, 
  {3(\Nf-4) \over (\Nf+1) (\Nf-3)} \bigr) \cr
\qp^{\a\adot}_1 &\sim \bigl( \fund,2; 1; {\Nf-1 \over \Nf-3}, 
  -\half{\Nf^2 - 3\Nf-2 \over \Nf-3}, 
  \half {(\Nf+3)(\Nf-4) \over (\Nf+1)(\Nf-3)} \bigr) \cr
\qp^{\a\adot}_2 &\sim \bigl( \fund,2; 1; {\Nf-1 \over \Nf-3}, 
  \half{(\Nf-2)(\Nf-1) \over \Nf-3},{3 \over 2} { (\Nf-1)(\Nf-4)\over 
  (\Nf+1)(\Nf-3)} \bigr) \cr
\qbar^+_\a &\sim \bigl( \antifund,1; 1; {(\Nf-4)(\Nf-1) \over \Nf-3}, 
  -{(\Nf-2)^2 \over \Nf-3}, - {\Nf^2-14\Nf+30 \over (\Nf+1)(\Nf-3)} \bigr) \cr
\qbar^3_\a &\sim \bigl( \antifund,1; 1; {(\Nf-4)(\Nf-1) \over \Nf-3}, 
  {\Nf-4 \over \Nf-3},  { 7\Nf-18 \over (\Nf+1)(\Nf-3)} \bigr) \cr
s_{\a\b} &\sim \bigl( \symbar,1; 1; -2{\Nf-1 \over \Nf-3}, -{2 \over \Nf-3}, 
 { 4\Nf-6 \over (\Nf+1)(\Nf-3)} \bigr) \cr
t^{\adot}_1 &\sim \bigl( 1,2; 1; 1-\Nf, -{\Nf+2 \over 2}, {1 \over 2}  
  {\Nf-4 \over \Nf+1} \bigr) \cr
t^{\adot}_2 &\sim \bigl( 1,2; 1; 1-\Nf, {\Nf-2 \over 2},  {3 \over 2}  
  {\Nf-4 \over \Nf+1} \bigr) \cr
m^{(ij)} &\sim \bigl(1,1; \sym;2, -2, 2 {\Nf-4 \over \Nf+1} \bigr) \cr
n^i_3 &\sim \bigl(1,1; \fund; 2-\Nf, -2,  2 {\Nf-4 \over \Nf+1} \bigr) \cr
n^i_- &\sim \bigl(1,1; \fund; 2-\Nf, \Nf-2,  3 {\Nf-4 \over \Nf+1} \bigr) \cr
n^\Nf_- &\sim \bigl(1,1; 1; 0, 2\Nf,   2 {\Nf-4 \over \Nf+1}\bigr) \cr}}
and tree level superpotential
\eqn\Wmagseven{\eqalign{
\Wmag & = m^{(ij)} q^\a_i s_{\a\b} q^\b_j 
 + n^i_- q^\a_i \qbar^+_\a 
 + n^i_3 q^\a_i \qbar^3_\a 
 + \e_{\adot\bdot} n^\Nf_- \qbar^+_\a \qp^{\a\adot}_1 t^{\bdot}_1 
 + \e_{\adot\bdot} {\qp}^{\a\adot}_1 s_{\a\b} {\qp}^{\b\bdot}_2 \cr
& \qquad  
 + \e_{\adot\bdot} \qbar^3_\a (\qp^{\a\adot}_1 t^{\bdot}_2 
   + \qp^{\a\adot}_2 t^{\bdot}_1) 
 + \e_{\adot\bdot} \qbar^+_\a \qp^{\a\adot}_2 t^{\bdot}_2.}}

	Only an $SU(\Nf-1) \times U(1)_\Nf$ subgroup of the global $SU(\Nf)$ 
that rotates the spinors in the $SO(7)$ theory is realized at short distance 
scales in this dual.  The subgroup's origin can be traced from the flows in 
\flowdiag.  One of the two 16-dimensional spinor fields is eaten when 
$SO(10)$ is Higgsed down to $SO(7)$, while the second breaks apart as $16 
\to 8+7+1$.  Nonrenormalizable terms pick out the lone $8$ from the $16$ 
and prevent its mixing with the $\Nf-1$ other spinors that come from
the $10$'s.  The full $SU(\Nf)$ symmetry is restored in the electric
theory only in the far infrared where all nonrenormalizable
interactions are negligible.  On the magnetic side, the $SU(\Nf)$ symmetry is 
realized in the infrared only at the quantum level.  Similar accidental 
restoration of global symmetries was first observed in $SU(2)$ duality 
\SeibergI\ and has been discussed in the recent literature 
\refs{\DistlerKarch,\LSII}.  

	As a final check on the $SU(\Nf-3) \times Sp(2)$ counterpart
to the $\NQ=2$ SO(10) model, we investigate its entry into the
confining phase when $\Nf=5$.  In this case, the dual gauge group
reduces to $\Gdual_\local = SU(2)_\L \times Sp(2)_\R$ where we have
appended left and right labels onto the two group factors in order to
distinguish them.  The $SU(2)_L$ and $Sp(2)_\R$ gauge groups are 
respectively infrared and asymptotically free at high energies.  
The latter grows strong at a scale $\LambdaR$ and confines the 
6 doublets $\qp$ and $t$ into the antisymmetric matrix \SeibergII\ 
\eqn\Mright{M_\R  = \pmatrix{ \qp^{\a\adot}_\I (\s_2)_{\adot\bdot} 
\qp^{\b\bdot}_\J & \qp^{\a\adot}_\I (\s_2)_{\adot\bdot} t^{\bdot}_\J \cr
t^{\bdot}_\J (\s_2^\T)_{\bdot\adot} \qp^{\a\adot}_\I  &
t^{\adot}_\I (\s_2)_{\adot\bdot} t^{\bdot}_\J \cr}.}
The mesons 
\eqn\sutworightinvariants{\eqalign{
k &= - (\s_2)_{\I\J} t^\adot_\I (\s_2)_{\adot \bdot} t^\bdot_\J \cr
(s_\R)^{(\a\b)} &= (\s_2)_{\I\J} \qp^{\a\adot}_\I (\s_2)_{\adot\bdot} 
  \qp^{\b\bdot}_\J \cr
(t_\R)^{[\a\b]}_\xrm &= i (\s_\xrm \s_2)_{\I\J} \qp^{\a\adot}_\I 
  (\s_2)_{\adot\bdot} \qp^{\b\bdot}_\J \cr
u_\R^\a &= (\s_2)_{\I\J} \qp^{\a\adot}_\I (\s_2)_{\adot\bdot} t^{\bdot}_\J \cr
(v_\R)^\a_\xrm &= i (\s_\xrm \s_2)_{\I\J} \qp^{\a\adot}_\I 
  (\s_2)_{\adot\bdot} t^\bdot_\J \cr }}
along with the $Sp(2)_\R$ invariants in \magmatter\ then represent the active 
matter degrees of freedom. Their dynamics are governed by the superpotential 
$W_\R = \Wtree + \WdynR$ where
\eqn\WtreeR{
\Wtree = m^{(ij)} q^\a_i s_{\a\b} q^\b_j + n^i_\xrm q^\a_i \qbar^\xrm_\a - 
s_{\a\b} (s_\R)^{\a\b} + \qbar^\xrm_\a (v_\R)^\a_\xrm }
and
\eqn\WRdyn{\eqalign{
\WdynR = {\Pf M_\R \over \LambdaR^3} &= {1 \over 8 \LambdaR^3} \Bigl\{
k (\Pf t_R^\xrm) (\Pf t_R^\xrm) - k \det s_\R
- u_\R^\T \s_2 s_\R \s_2 u_\R - (v_\R)^\T_\xrm \s_2 s_\R \s_2 (v_\R)_\xrm \cr
& \qquad
- \e_{\xrm \yrm \zrm} (v_\R)^\T_\xrm \s_2 (t_\R)_\yrm \s_2 (v_\R)_\zrm 
 + (v_\R)^\T_\xrm \s_2 (t_\R)_\xrm \s_2 u_\R
 - u_\R^\T \s_2 (t_\R)_\xrm \s_2 (v_\R)_\xrm \Bigr\}. }}

	The last two terms in \WtreeR\ render massive the $SU(2)_\L$
triplets $s$ and $s_\R$ as well as the doublets $\qbar$ and $v_\R$.
Once these heavy fields are integrated out, only six $SU(2)_\L$ doublets 
in $q$ and $u$ remain.  Below a scale $\LambdaL$, the $SU(2)_\L$ color 
force confines these fields into the meson matrix
\eqn\Mleft{M_\L = \pmatrix{ q^\a_i (\s_2)_{\a\b} q^\b_j &
q^\a_i (\s_2)_{\a\b} u_\R^\b \cr
u_\R^\b (\s_2^\T)_{\b\a} q^\a_i & 0 \cr}}
and generates the additional superpotential term $\WdynL = 
\Pf M_\L / \LambdaL^4$.  After collecting together the separate quantum and
classical contributions and renaming all composites in terms of the $\Nf=5$
magnetic baryons 
\eqn\magbaryonsfive{\eqalign{
p_{ij} &= \e_{\a\b} q^\a_i q^\b_j \cr
r_i &= \e_{\a\b} \e_{\adot\bdot} \e_{\I\J} q^\a_i q^{\b\bdot}_\I t^\adot_\J = 
  - \e_{\a\b} q^\a_i (u_\R)^\b \cr
t_\xrm &= \e_{\a\b} \e_{\adot\bdot} (\s_\xrm \s_2)_{\I\J} \qp^{\a\adot}_\I 
\qp^{\b\bdot}_\J = i \e_{\a\b} (t_\R)^{\a\b}_\xrm, \cr}}
we obtain the  superpotential which controls the dynamics in the 
extreme infrared:
\eqn\Wtot{\eqalign{W_{\rm tot} &= \Wtree+\WdynL+\WdynR \cr
&\simeq k t_\xrm t_\xrm 
  + \e_{\xrm \yrm \zrm} n^i_\xrm n^j_\yrm p_{ij} t_\zrm 
  + n^i_\xrm r_i t_\xrm + \e^{ijklm} p_{ij} p_{kl} r_m \cr
& \qquad + r_i m^{ij} r_j 
  + k m^{ij} m^{kl} p_{ik} p_{jl}
  + m^{ij} n^k_\xrm n^l_\xrm p_{ik} p_{jl}.}}
The functional form of this magnetic result coincides with the first seven
terms in the electric superpotential in \Wconfining.  As in all previous
Pouliot-type duals, there are some remaining nonrenormalizable terms whose 
origin we have not been able to identify.  But aside from these last four 
terms, we see that the magnetic theory properly reproduces the confining phase 
for the $\NQ=2$ $SO(10)$ model.

\newsec{The general dual}

	We now consider the dual to the $SO(10)$ theory with arbitrary 
numbers of spinors and vectors.  The enlarged electric theory has symmetry 
group 	
\eqn\newsymgroup{G = SO(10)_{\rm local} \times \bigl[ SU(\Nf) \times SU(\NQ) 
\times U(1)_\Y \times U(1)_\R \bigr]_{\rm global}}
and superfield matter content 
\foot{To avoid excessive cluttering of composite operator indices, we 
label $SO(10)$ spinors with lower flavor indices even though they transform 
according to the fundamental irrep of $SU(\NQ)$.}
\eqn\newmatter{\eqalign{
V^i_\mu & \sim \bigl( 10; \fund ,1 ; -2\NQ, R \bigr) \cr
Q^\A_\I  & \sim \bigl( 16; 1,\fund; \Nf, R \bigr) \cr}}
where $R = 1-8/(\Nf+2\NQ)$.  Various operators act as gauge invariant 
coordinates on the moduli space of this new theory.  We will focus upon 
those which generalize the $\NQ=2$ composites in \eleccomposites:
\eqn\neweleccomposites{\eqalign{
K_{\I\J\K\L} &= Q^\T_{(\I} \Gamma^\mu C Q_{\J)} Q^\T_{(\K} \Gamma_\mu C Q_{\L)}
\sim \bigl(1; 1, \twotwo; 4\Nf, 4R \bigr) \cr
M^{(ij)} &= (V^\T)^{i \mu} V^j_\mu 
	\sim \bigl(1; \sym, 1; -4\NQ, 2 R \bigr) \cr
N^i_{(\I\J)} &= Q^\T_{(\I} \Vslash^i C Q_{\J)} 
	\sim \bigl(1; \fund,\sym; 2 \Nf - 2\NQ, 3 R \bigr) \cr
P^{[ijk]}_{[\I\J]} &= {1 \over 3!} Q^\T_{[\I} \Vslash^{[i} \Vslash^j 
  \Vslash^{k]} C Q_{\J]}
  \sim \bigl(1; \antithree,\anti; 2\Nf - 6 \NQ, 5 R \bigr) \cr 
R^{[ijkl]}_{\I\J\K\L} &= {1 \over 4!} Q^\T_{(\I} \Gamma^\mu Q_{\J)} 
  Q^\T_{(\K} \Gamma_\mu \Vslash^{[i} \Vslash^j \Vslash^k \Vslash^{l]} C Q_{\L]}
  \sim \bigl(1; \antifour,\twotwo; 4\Nf - 8 \NQ, 8 R \bigr) \cr 
T^{[ijklm]}_{(\I\J)} &= {1 \over 5!} Q^\T_{(\I} \Vslash^{[i} \Vslash^j 
  \Vslash^k \Vslash^l \Vslash^{m]} C Q_{\J)}
  \sim \bigl(1; \antifive,\sym; 2\Nf - 10 \NQ, 7 R \bigr). \cr }}
Determining the quantum numbers for the two-spinor operators $M$, $N$, $P$ and 
$T$ is straightforward.  On the other hand, figuring out the $SU(\NQ)$ irrep 
assignments for the four-spinor hadrons $K$ and $R$ is not so trivial.
Previously when we had only $\NQ=2$ spinor flavors, we used counting and 
anomaly arguments to deduce that these baryons were global $SU(2)$ singlets.  
Now that we have expanded the electric theory to include arbitrary numbers of 
spinors, neither the number of independent quartic spinor $SO(10)$ invariants 
nor their transformation rules under $SU(\NQ)$ are immediately obvious.

	In order to address these questions, it is useful to recall two 
Fierz identities \Kennedy:
\eqn\fierz{\eqalign{
& \Bigl(\Gamma^{[\mu} \Gamma^\nu \Gamma^\s \Gamma^\t \Gamma^{\lambda]} 
  P_\pm \Bigr)_{\A\B} 
\Bigl(\Gamma_{[\mu} \Gamma_\nu \Gamma_\s \Gamma_\t \Gamma_{\lambda]} 
  P_\pm \Bigr)_{\C\D} =0 \cr
& \Bigl(\Gamma^{[\mu} \Gamma^\nu \Gamma^{\lambda]} P_\pm \Bigr)_{\A\B} 
\Bigl(\Gamma_{[\mu} \Gamma_\nu \Gamma_{\lambda]} P_\pm \Bigr)_{\C\D} =
12 \Bigl(\Gamma^\mu P_\pm \Bigr)_{\A\B} \Bigl(\Gamma^\mu P_\pm \Bigr)_{\C\D}
+ 24\Bigl(\Gamma^\mu P_\pm \Bigr)_{\A\D} \Bigl(\Gamma^\mu P_\pm 
\Bigr)_{\C\B}. \cr}}
The first relation implies that the $SO(10)$ invariant
$Q_{(\I} \Gamma^{[\mu \nu \s\t\lambda]} C Q_{\J)} 
 Q_{(\K} \Gamma_{[\mu \nu \s\t\lambda]} C Q_{\L)}$ simply vanishes.
The second guarantees that the operator
$Q_{[\I} \Gamma^{[\mu \nu \lambda]} C Q_{\J]} Q_{[\K} 
\Gamma_{[\mu \nu \lambda]} C Q_{\L]}$ can be decomposed in terms of others 
of the form $Q_{(\I} \Gamma^\mu C Q_{\J)} Q_{(\K} \Gamma_\mu C
Q_{\L)}$.  So without loss of generality, we need only consider
four-spinor composites of this last type.  Such hadrons transform
under $SU(\NQ)$ according to the product representation
\eqn\productrep{\Bigl(\sym \times \sym \Bigr)_\S = \four + \twotwo.}
Since the $SO(10)$ theory with just one spinor has no flat directions \ADSII, 
holomorphic invariants associated with the totally symmetric irrep $\four$ 
must vanish.  We thus deduce that four-spinor operators transform only 
according to the ``window frame'' irrep $\twotwo$.

	The expanded $SO(10)$ model is infrared free when its
Wilsonian beta function coefficient $b_0 = 24 - \Nf - 2 \NQ$ is
nonpositive.  In contrast, the theories with $\NQ=1$, 2, 3 and 4 spinors
confine provided they respectively possess $\Nf\leq 6$, 5, 3 and 1 vectors. 
In between these two limits, the $SO(10)$ model is asymptotically free, but
gauge singlet operators do not suffice to describe its infrared behavior. 
Instead, the physics of this last phase can be described in terms of a dual 
which naturally generalizes our earlier $SU(\Nf-3) \times Sp(2)$ product
theory.  

	The magnetic dual is based upon the symmetry group
\eqn\newmaggroup{\Gdual = \bigl[SU(\Ncdual) \times Sp(2\Ncpdual) \bigr]_{\rm 
local} \times \bigl[ SU(\Nf) \times SU(2) \times U(1)_\Y \times U(1)_\R 
\bigr]_{\rm global}}
where $\Ncdual = \Nf+2\NQ-7$ and $\Ncpdual = \NQ-1$.  It is important
to note that only an $SU(2)$ subgroup of the global $SU(\NQ)$ which
rotates the spinors in the electric theory is realized in the
ultraviolet on the magnetic side.  The $SU(2)$ subgroup is embedded inside 
$SU(\NQ)$ such that the fundamental $\NQ$-dimensional irrep of the latter maps 
onto the $\NQ$-dimensional irrep of the former.  The full $SU(\NQ)$ global 
symmetry is realized in the magnetic theory only at long distances.

	The new dual's matter content generalizes that which we previously 
found in \magmatter:
\eqn\newmagmatter{\eqalign{
q^\a_i &\sim \bigl( \fund ,1 ; \antifund,1; Y_q, R_q \bigr) \cr
\qp^{\a\adot}_\I &\sim \bigl( \fund,\fund; 1,2; Y_{q'}, R_{q'} \bigr) \cr
\qbar_{\a (\I_1 \cdots \I_{2\NQ-2})}  &\sim \bigl( \antifund,1; 1,2\NQ-1; 
  Y_\qbar, R_\qbar \bigr) \cr
s_{\a\b} &\sim \bigl( \symbar,1; 1,1; Y_s, R_s \bigr) \cr
t^{\adot}_{(\I_1 \cdots \I_{2\NQ-3})} &\sim \bigl( 1,\fund; 1,2\NQ-2; 
  Y_t, R_t \bigr).  \cr
m^{(ij)} &\sim \bigl(1,1; \sym,1; Y_m, R_m \bigr) \cr
n^i_{(\I_1 \cdots \I_{2\NQ -2})} &\sim \bigl(1,1; \fund,2\NQ-1; 
         Y_n, R_n \bigr). \cr}}
The hypercharge and R-charge assignments for these fields are listed below:
\eqn\charges{\eqalign{
Y_q & = {2 \over \Ncdual} (\NQ \Ncdual - \Nf \Ncpdual) \cr
Y_{q'} & = -2\Nf{\Ncpdual \over \Ncdual} \cr
Y_\qbar & =   -2\Nf{\Ncdual - \Ncpdual \over \Ncdual}  \cr
Y_s & = 4\Nf{\Ncpdual \over \Ncdual}  \cr
Y_t & = 2 \Nf   \cr 
Y_m & = -4\NQ   \cr
Y_n & = 2\Nf-2\NQ \cr}
\qquad\qquad
\eqalign{
R_q &= {9 - 2 \NQ \over \Ncdual} R   \cr
R_{q'} &= {\Nf+2 \over \Ncdual} R   \cr
R_\qbar &=   -{\Nf^2+2\Nf\NQ-22\Nf-28\NQ+96 \over \Ncdual(\Nf+2\NQ)}  \cr
R_s &= 2{4\NQ^2+2\Nf\NQ-\Nf-18 \NQ+16 \over \Ncdual(\Nf+2\NQ)}   \cr
R_t &= 2R.  \cr
R_m &= 2R  \cr
R_n &= 3R. \cr}}
The tree level magnetic superpotential similarly extends \Wmagnetic\ 
in the $\NQ=2$ model:
\eqn\newWmagnetic{\eqalign{
\Wmag &= {1 \over \mu_1^2} m^{(ij)} q^\a_i s_{\a\b} q^\b_j 
 + \lambda_1 \e^{\I\J} J_{\adot \bdot} {\qp}^{\a\adot}_\I s_{\a\b} 
  {\qp}^{\b\bdot}_\J  \cr
& \quad  + {1 \over \mu_2^2} \e^{\I_1 \J_1} \cdots \e^{\I_{2\NQ-2} 
  \J_{2\NQ-2}} \>
   q^\a_i n^i_{(\I_1 \cdots \I_{2\NQ-2})} \qbar_{\a (\J_1 \cdots \J_{2\NQ-2})} 
  \cr
& \quad  
 + \lambda_2 \e^{\I_1 \J_1} \cdots \e^{\I_{2\NQ-2} \J_{2\NQ-2}} 
  J_{\adot \bdot} \>
  {\qp}^{\a\adot}_{\J_{2\NQ-2}}  \qbar_{\a (\I_1 \cdots \I_{2\NQ-2})} 
   t^{\bdot}_{(\J_1 \cdots \J_{2\NQ-3})}. \cr}}
Here $J_{\adot \bdot} = \bigl(1_{\Ncpdual \times \Ncpdual} \otimes 
i\s_2\bigr)_{\adot \bdot}$ denotes the antisymmetric metric that remains 
invariant under $Sp(2\Ncpdual)$ rotations.  The coefficients $\mu_{1,2}$ and 
$\lambda_{1,2}$ represent dimensionful and dimensionless coupling constants
which we again set to unity for simplicity.

\midinsert
\parasize=1in

\begintable
Global Anomaly \| Value \crthick
$SU(\Nf)^3$ 		\| $10$ \nr
$SU(\Nf)^2 U(1)_\Y$	\| $-20 \NQ$ \nr
$SU(\Nf)^2 U(1)_\R$	\| $-\displaystyle {80 \over \Nf+2\NQ}$ \nr
$SU(2)^2 U(1)_\Y$	\| $\displaystyle {8 \over 3} \Nf \NQ (\NQ^2-1)$ \nr
$SU(2)^2 U(1)_\R$	\| $\displaystyle {64 \NQ (1-\NQ^2) \over 
			   3(\Nf+2\NQ)} $ \nr
$U(1)_\Y$		\| $-4 \Nf \NQ$ \nr
$U(1)^3_\Y$		\| $16 \Nf \NQ (\Nf^2 - 5 \NQ^2)$ \nr
$U(1)_\R$		\| $-\displaystyle{ 38 \NQ + 35 \Nf \over \Nf+2\NQ}$\nr
$U(1)^3_\R$		\| $\displaystyle{{\rm numer} \over (\Nf+2\NQ)^3 }$ \nr
$U(1)^2_\Y U(1)_\R$	\| $-64 \Nf \NQ \displaystyle{2 \Nf+5 \NQ \over 
				\Nf+2\NQ}$ \nr
$U(1)^2_\R U(1)_\Y$	\| $- 256 \displaystyle{\Nf\NQ \over (\Nf+2\NQ)^2}$ 
\endtable
\bigskip
\parasize=5in
\centerline{\para{Table 3: Global `t~Hooft anomalies in the $SO(10)$ theory 
with $\Nf$ vectors and $\NQ$ spinors and its $SU(\Ncdual) \times Sp(2\Ncpdual)$ 
dual.  The numerator of the $U(1)^3_\R$ anomaly equals $360 \NQ^3 + 45 \Nf^3 
+ 540 \NQ^2 \Nf + 270 \Nf^2 \NQ - 8192 \NQ - 5120 \Nf$.}}
\bigskip
\endinsert

	Several points about this new product dual should be noted.
Firstly, it satisfies all necessary anomaly checks.  The
$SU(\Ncdual)^3$, $SU(\Ncdual)^2 U(1)_\Y$, $SU(\Ncdual)^2 U(1)_\R$,
$Sp(2\Ncpdual)^2 U(1)_\Y$, and $Sp(2\Ncpdual)^2 U(1)_\R$ anomalies
vanish, and an even number of fundamentals transform under the
$Sp(2\Ncpdual)$ gauge group.  All anomaly matching conditions
associated with the common $SU(\Nf) \times SU(2) \times U(1)_\Y \times
U(1)_\R$ global symmetry group are also satisfied.  We display these
`t~Hooft anomaly values in Table~3.  Secondly, we observe from the Wilsonian
beta functions coefficients $\tilde{b_0}^{SU(\Ncdual)}=2
\Nf + 2 \NQ - 16$ and $\tilde{b_0}^{Sp(2\Ncpdual)}= 8 - \Nf$ that the dual 
never resides within a free magnetic phase for any values of $\Nf$ and
$\NQ$.  We also note that the $\NQ=3$ ($\NQ=4$) magnetic theory enters
into the confining phase when $\Nf =3$ ($\Nf=1$) in a fashion similar
to that for the $\NQ=2$ model.  Finally, the symplectic gauge group in 
\newmaggroup\ sees only fundamental irrep matter. By dualizing the
$Sp(2\Ncpdual)$ color factor, we can derive another magnetic
counterpart to the general $SO(10)$ model 
\refs{\SeibergI,\IntriligatorPouliot}.  The symmetric tensor disappears from 
the resulting product dual, and three antisymmetric fields take its place.  
We outline the basic structure of this second dual in Appendix~B.  As it is 
more complicated but not more illuminating than the magnetic theory we have 
already discussed, we will not consider it further.

	As an additional check, we consider flows induced by 
mass deformations.  If we add a tree level superpotential 
$W = m V^{\Nf}V^{\Nf}$ which gives mass to a vector in the 
$SO(10)$ theory, we find that the magnetic $SU(\Nf-2\NQ-7)$ 
color group properly breaks down to $SU([\Nf-1]-2\NQ-7)$.   On the other hand, 
we cannot give mass to any of the $SO(10)$ spinors.  Instead, we may Higgs 
$SO(10)$ down to $SO(9)$ along the flat direction with nonzero 
$\vev{V^{\Nf}}$, and then add a mass term $W =\vev{V^{\Nf}}Q_\NQ Q_\NQ$ which 
removes one spinor.  The effect on the magnetic theory is quite complicated.  
A long and careful analysis demonstrates that the gauge group is properly 
reduced and the correct matter fields are removed.  In the case when $\NQ=2$, 
the $SU(\Nf-3)\times Sp(2)$ gauge group breaks to $SU(\Nf-5)$, and instanton 
effects in the broken $Sp(2)$ generate a $\det s$ term in the magnetic 
superpotential.  The resulting theory thus reproduces 
the dual counterpart of $SO(9)$ with one spinor, which follows from the 
dual to the $\NQ=1$ $SO(10)$ model \refs{\PouliotStrasslerII,\ChoI}.  Since 
all earlier Pouliot-type duals can be obtained from the $\NQ=1$ theory, our 
duality transformation contains these results as special cases.

	We next examine the mapping of gauge invariant operators between the 
electric and magnetic theories.  This issue is complicated by the partial 
$SU(2)$ realization of the $SU(\NQ)$ global symmetry in the dual.  In order to 
match operators, we need to first decompose the electric hadrons' 
$SU(\NQ)$ representations under the $SU(2)$ subgroup following the embedding 
$\fund \to \NQ$.  For example, symmetric and antisymmetric tensors of 
$SU(\NQ)$ break apart as
\eqn\irrepdecomp{\eqalign{
\sym    \quad & \to \quad \sum_{i=0}^\imax \bigl( 2\NQ-1-4i 
\bigr) \qquad {\rm where} \sp \imax = \cases{(\NQ-1)/2 & $\NQ$ odd \cr
					     (\NQ-2)/2 & $\NQ$ even \cr} \cr
\anti    \quad & \to \quad \sum_{i=0}^\imax \bigl( 2\NQ-3-4i 
\bigr) \qquad {\rm where} \sp \imax = \cases{(\NQ-3)/2 & $\NQ$ odd \cr
					    (\NQ-2)/2 & $\NQ$ even. \cr} \cr}}
Combining this information with abelian charge assignments, we 
readily find that the dual baryons
\eqn\nptbaryons{\eqalign{
p &= q^{\Nf-3} \qp^{2\NQ-4} \cr
t &= q^{\Nf-5} \qp^{2\NQ-2} \cr}}
match onto $P$ and $T$ in \neweleccomposites.  Similarly, the elementary $n$ 
partons in \newmagmatter\ along with the magnetic composites
\eqn\nbaryons{n' = q^{\Nf-1} \qp^{2\NQ-6}}
account for all the $N$ fields in the electric theory.

	Whereas mapping two-spinor $SO(10)$ operators is straightforward, 
finding dual counterparts to four-spinor hadrons is much more involved.  
We will concentrate upon identifying the magnetic partners to the operator $K$ 
in \neweleccomposites.  First, we decompose its $SU(\NQ)$ $\twotwo$ irrep 
under the $SU(2)$ subgroup.  The results for small values of $\NQ$ are 
displayed in Table~4.  We next observe that the composites $t^2$, 
$q^\Nf \qp^{\Ncdual-\Nf} t$ and $(q^\Nf \qp^{\Ncdual-\Nf})^2$ have 
the same hypercharge and R-charge as $K$.  The total number of these magnetic 
hadrons naively appears to exceed ${\rm dim}(\twotwo \hskip-0.5 em ) = 
\NQ^2 (\NQ^2-1)/12$.  But it is important to remember that we should only 
count operators which are 
not set to zero by equations of motion.  After carefully considering the 
effect of each term in $\Wmag$, we find that the numbers of such nonredundant 
$t^2$, $q^\Nf \qp^{\Ncdual-\Nf} t$ and $(q^\Nf \qp^{\Ncdual-\Nf})^2$ baryons 
respectively equal $\nobreak{(2\NQ-3)(\NQ-1)}$, $(2\NQ-3)(\NQ-2)(\NQ-3)/2$ and 
$(\NQ-2)(\NQ-3)^2(\NQ-4)/12$.  As Table~4 illustrates, these magnetic 
composites precisely account for all components of $K$ in the electric theory. 

\midinsert
\parasize=1in

\begintable
$\quad \NQ \quad $ \| $ \quad K \quad $ \| $\quad t^2 \quad $ 
\| $q^\Nf \qp^{\Ncdual-\Nf} t$ 
\| $(q^\Nf \qp^{\Ncdual-\Nf})^2$ \crthick
2 \| 1=(1)  \| 1 \| 0 \| 0 \cr
3 \| 6=(5)+(1)  \| 6 \| 0 \| 0 \cr
4 \| 20=(9)+2(5)+(1) \| 15 \| 5 \| 0 \cr
5 \| 50=(13)+2(9)+(7)+2(5)+2(1) \| 28 \| 21 \| 1 \endtable
\bigskip
\parasize=5in
\centerline{\para{Table 4:  Mapping of the electric theory four-spinor 
composite $K$ onto nonredundant magnetic baryons.  The $SU(2)$ decomposition 
of the $SU(\NQ)$ $\twotwo$ irrep is displayed in the second column.}}
\endinsert

	The final check of the duality transformation which we perform 
involves an intricate renormalization group flow analysis that generalizes 
the procedure introduced in ref.~\PouliotStrasslerI.  We first deconfine the 
magnetic theory's symmetric tensor and then consider the resulting 
$SO(\tilde\Nc+4)\times SU(\tilde\Nc)\times Sp(2\tilde\Nc')$ model.  
We investigate flows in this theory for large and small values of the 
ratio $\Lambda_{SO}/\Lambda_{SU}$, where $\Lambda_{SO}$ $[\Lambda_{SU}]$ 
denotes the strong coupling scale for $SO(\Ncdual+4)$ $[SU(\Ncdual)]$.
Since holomorphy ensures that phase transitions between these two regions 
cannot occur \SeibergII, the low energy physics of both must be qualitatively 
similar.  After utilizing several different duality transformations, 
we find that the triple product theory flows to the same $SO(10)$ fixed point 
for both large and small values of $\Lambda_{SO}/\Lambda_{SU}$.  This 
demonstrates that our dual is consistent with Seiberg's well-known 
results.

	We first consider the $\Lambda_{SO}\gg\Lambda_{SU}$ case.  The 
$SO(\Ncdual+4)$ factor then grows strongly coupled, and the renormalization 
group flow passes through the intermediate stages 

\bigskip
\centerline{$SO(N_f+2N_Q-3)\times SU(N_f+2N_Q-7)\times Sp(2\NQ-2)$}
\medskip
\hskip 2.35 in {$\downarrow$ (SO confinement)}
\medskip
\vskip -0.2 in
\eqn\flowone{SU(\Nf+2\NQ-7)\times Sp(2N_Q-2)}
\medskip
\hskip 2.35 in {$\downarrow$ (SU $\times$ Sp duality)} 
\medskip
\centerline{$SO(10)$}
\noindent
before arriving at the final infrared fixed point.  On the other hand, the 
flow pattern takes the form

\bigskip
\centerline{$SO(\Nf+2\NQ-3) \times SU(\Nf+2\NQ-7) \times Sp(2\NQ-2)$}
\medskip
\hskip 2.35 in {$\downarrow$ (SU duality)}
\medskip
\centerline{$SO(\Nf+2\NQ-3) \times SU(2\NQ+3) \times Sp(2\NQ-2)$}
\medskip
\hskip 2.35 in {$\downarrow$ (SO duality)}
\medskip
\vskip -0.2 in
\eqn\flowtwo{SO(10) \times SU(2\NQ+3) \times Sp(2 \NQ-2)}
\medskip
\hskip 2.35 in {$\downarrow$ (SU $\times$ Sp duality)}
\medskip
\centerline{$SO(10) \times SO(10)$}
\medskip
\hskip 2.35 in {$\downarrow$ (Tree level breaking)}
\medskip
\centerline{$SO(10)$}
\noindent
when $\Lambda_{SU} \gg \Lambda_{SO}$.  The first two steps involve
Seiberg's duality transformations \refs{\SeibergI,\IntriligatorSeiberg}, 
while the third utilizes the variant of our duality transformation
discussed in Appendix~C.  The closure of the two duality chains in \flowone\ 
and \flowtwo\ constitutes a highly nontrivial consistency check on our 
results.

	The details underlying this flow analysis are presented in 
Appendix~D.  The interested reader will find them remarkable in their 
complexity and intricacy.

\newsec{A composite Standard Model}

	In his original work on $\CN=1$ duality, Seiberg speculated
that the Standard Model might represent a low energy effective
description of a more fundamental theory with a totally different
gauge group \SeibergI.  In such a scenario, some or all of the
ordinary matter and gauge particles would be composite.  As the
Minimal Supersymmetric Standard Model is weakly coupled at all
energies between the SUSY breaking and Planck scales, there is little
point in constructing a dual to it.  But if Nature contains additional
vectorlike matter with masses between $\Lambda_{\rm SUSY}$ and 
$\Lambda_{\rm Planck}$, the Standard Model with these extra fields could 
become strongly interacting at high energies.  In this case, it would clearly
be beneficial to find a weakly coupled description of the microscopic
physics.

	If one simply adds matter charged in fundamental
representations and tries to apply Seiberg's $SU(N)$ duality
transformation to the individual nonabelian factors of the Standard
Model gauge group, one encounters an unending sequence of duals
involving ever larger gauge groups and shorter energy regimes 
\MJSSCGT.  Some attempts to evade this ``Duality Wall'' have
been made in the past \Maekawa.  But to properly overcome this problem, one
needs either to dualize two or three of the Standard Model's
subgroups simultaneously or to unify the Standard Model within a
larger group.  We now present a toy model which follows the second approach.

	We start with an $SO(10)$ grand unified theory. 
In order to break $SO(10)$ down to $SU(3) \times SU(2) \times U(1)$, we need 
a scalar field in the adjoint representation.  Since we do not yet know a 
simple dual to an $SO(10)$ theory with spinor, vector and adjoint matter, we 
adopt the deconfinement method to build the adjoint \Berkooz.  
$Sp(2\Nc)$ theories with $2N_c+4$ fields in the fundamental representation
confine and are described at low energy in terms of mesons 
antisymmetric in flavor.  Consequently, if we take an $SO(10)$ theory
with $N_f$ vectors and gauge an $Sp(6)$ subgroup of its $SU(\Nf)$
flavor symmetry, we can generate an adjoint of $SO(10)$ below the
$Sp(6)$ confining scale.

	We are thus motivated to consider a product theory with nonabelian 
symmetry group $ \bigl[ SO(10) \times Sp(6) \bigr]_{\rm local} \times
\bigl[ SU(\Nf) \times SU(3) \bigr]_{\rm global}$
and matter content $V^i_\mu  \sim (10,1; \Nf,1)$, $Z^\adot_\mu  \sim
(10,6; 1,1)$, $Q^\A_\I  \sim (16,1; 1,3)$. The $SO(10)$ factor is
free in the infrared provided $\Nf \ge 12$.  On the other hand, the
$Sp(6)$ gauge group grows strong at energies below its strong coupling
scale $\Lambda_{SP}$.  It confines the $Z$ partons into the
antisymmetric mesons $A_{[\mu \nu]} = Z^\adot_\mu J_{\adot \bdot}
Z^\bdot_\nu$ and dynamically generates the superpotential $\Wdyn = \Pf
A / \Lambda_{SP}^7$.  When this quantum effect is combined with an 
appropriate classical superpotential $W=f(A)$, the $SO(10)$ gauge
group breaks to  $ SU(3)_\C \times
SU(2)_\L \times U(1)_\Y \times U(1)_{\B-\L}$.  The resulting low
energy theory then possesses the Standard Model gauge group and chiral
matter content along with extra vectorlike matter fields.  These
ingredients are precisely those we need to investigate Seiberg's
suggestion.

	Using the results of section~4, we regard this $SO(10)
\times Sp(6)$ theory itself as the low energy limit of an $SU(\Nf+5)
\times Sp(4) \times Sp(6)$ model, provided $\Nf \ge 12$.  The $Sp(6)$
factor acts as a weakly coupled spectator under the $SO(10)
\leftrightarrow SU(\Nf-5) \times Sp(4)$ duality transformation. 

	The principle features of our toy model are thus the following.
At the Planck or string scale, it is based upon an $SU(\Nf+5) \times Sp(4) 
\times Sp(6)$ gauge group.  This theory flows down to an $SO(10) \times Sp(6)$ 
model.  The $SO(10)$ factor is infrared free for $\Nf\geq 12$, while the 
$Sp(6)$ force grows strong at low energies and confines six vectors into an
$SO(10)$ adjoint.  A tree level superpotential can induce an
expectation value for the adjoint field which breaks the intermediate
grand unified theory down to the Standard Model.  Since the adjoint
and extra vectors are nonchiral, we can arrange for unwanted
components of these fields to develop large masses.  We are then left
with three generations of Standard Model families at low energies.
Additional terms in the classical superpotential may be added to
provide Yukawa couplings to these matter fields.

	This model is certainly contrived.  However, it has a sensible high 
energy description and flows to the Standard Model at low energies as a 
consequence of strong coupling effects.  If some scenario like this actually 
operates in the real world, all Standard Model gauge bosons and matter fields
could indeed represent composite low energy degrees of freedom.

\newsec{Conclusion}

	The duals which we have constructed in this article exhibit a
number of novel features.  Firstly, our transformation represents an
essentially new type of product group duality.  It does not follow 
from Seiberg's results, and it is not related by confinement to other known 
dual pairs.  Secondly, it provides dual descriptions of supersymmetric 
theories with arbitrary numbers of two distinct types of matter.  All 
previous electric theories for which magnetic duals have been found involve at
most one infinite chain of matter fields.  As can be seen in \flowdiag, 
counterparts to numerous other theories may be derived along various spinor 
and vector flat directions.  Thirdly, the role which accidental symmetries 
plays in these theories is unusual.  The $SO(10)$ model with $\Nf$ vectors and 
$\NQ$ spinors possesses an $SU(\NQ)$ global symmetry.  Only an $SU(2)$ 
subgroup, under which the spinors transform according to the $\NQ$-dimensional 
representation, is manifest in the $SU(\Ncdual) \times Sp(2\Ncpdual)$ dual.
As a result, the size of the global subgroup present in the classical dual 
does not depend upon $\NQ$.  Instead, the $\NQ$ dependence resides in the 
matter representations of the dual theory.  Finally, $SO(10)$ models with 
spinors have clear implications for particle physics.   We have constructed a 
toy model with a completely unfamiliar gauge group and matter content which 
flows down to the Standard Model via a three-generation $SO(10)$ grand unified 
theory.  This scenario realizes Seiberg's suggestion that all Standard Model 
gauge bosons and matter fields might represent low energy effective degrees of 
freedom of some more fundamental gauge theory.

\bigskip\bigskip\bigskip\bigskip
\centerline{{\bf Acknowledgments}}
\bigskip

	We thank Howard Georgi for helpful discussions.  We also 
acknowledge support from  DOE Grant DE-FG02-96ER40559 (MB), the National 
Science Foundation under Grant \#PHY-9218167 (PC), DOE grant 
DE-FG03-92-ER40701 and the DuBridge Fellowship Foundation (PK), and the 
National Science Foundation under Grant PHY-9513835 and the WM Keck Foundation 
(MJS).  

\appendix{A}{Mapping $\pmb{Q^4 W^2}$}

	In $SO(\Nc)$ dual pairs, gauge invariant operators involving the field 
strength $W$ play an important role in the matching of chiral rings 
\IntriligatorSeiberg.  We have already seen such composites in \elecbaryons\ 
- \magexotics.  One can construct many more of this type.  In this appendix, we 
examine the mapping of primary $Q^4 W^2$ operators in the $\NQ=2$ $SO(10)$ 
model.  We adopt the gauge conventions of refs.~\refs{\Wess{--}\WessBagger}\ 
and let ${\cal D}$ denote a super convariant derivative which includes the 
gauge connection.

	$W^2 = \e^{ab} W_a W_b$ transforms under $SO(10)$ according to 
the symmetric product $(45 \times 45)_\S =1+54+210+770$.  We recall from 
\spinorprod\ that $Q^2$ transforms as $16 \times 16=10_\S +120_\A +126_\S$.  
The relevant $Q^4$ products we need to consider in order to form $SO(10)$ 
invariants are thus $10 \times 10=54+...$, $10 \times 120=210+...$, 
$10 \times 126=210+...$, $120 \times 120=54+210+...$, $120 \times 126=210+...$ 
and $126 \times 126=54+... \> $.  Most of the possible $Q^4 W^2$ combinations 
are either not primary or else not independent.  For example, the 
$10 \times 10$ operator $\e_{ab} Q \Gamma^\mu Q Q\Gamma^\nu Q
W^a_{\mu\sigma} W^b_{\nu\sigma}$ is a descendant as
$W^b_{\sigma\nu} Q \Gamma^\nu Q \propto 
\bar{\cal D}_{\dot a} \bar{\cal D}^{\dot a} {\cal D}^b  
Q \Gamma_\sigma Q$ \BerkoozII. Similarly, $Q \Gamma^{[\mu_1..\mu_l]} 
\Gamma^{[\nu_1 \nu_2]} Q W_{\nu_1\nu_2}$ is a descendant since 
$W^b_{\nu_1 \nu_2} \Gamma^{\nu_1 \nu_2} Q = 
\bar{\cal D}_{\dot a} \bar{\cal D}^{\dot a} {\cal D}^b Q$.  As a result, 
$Q^4$ products arising from $10 \times 120$ yield only descendant $Q^4 W^2$ 
operators.

	After a systematic search, we find just two independent primaries which 
come from the $120 \times 120$ product:
\eqn\primaries{\eqalign{
U &= \e_{ab} Q_1 \Gamma^{[\mu_1\mu_2\mu_3]} Q_2 
Q_1 \Gamma^{[\mu_1\nu_2\nu_3]} Q_2 W^a_{\mu_2\nu_2} W^b_{\mu_3\nu_3} \cr
V &= \e_{ab} Q_1 \Gamma^{[\mu_1\mu_2\mu_3]} Q_2 
Q_1 \Gamma^{[\mu_1\mu_2\nu_3]} Q_2 W^a_{\mu_3\sigma} W^b_{\nu_3\sigma}. 
\cr}}
Both of these operators are singlets under the global $SU(2)$ which rotates the 
two spinors in the electric theory.  On the magnetic side, 
\eqn\magprimaries{\eqalign{
u & = \det s \cr
v & = \e_{\adot\bdot} \e_{\cdot \ddot} \e^{\I\J} \e^{\K\L} 
\qp^{\a\adot}_\I t^{\bdot}_\J s_{\a\b} \qp^{\b\cdot}_\K t^{\ddot}_\L \cr}}
share precisely the same quantum numbers as $U$ and $V$.  We thus
identify the electric operators with linear combinations of the magnetic 
operators.

\appendix{B}{A second general dual} 

	It is possible to construct another dual to the $SO(10)$ model with 
$\Nf$ vectors and $\NQ$ spinors which differs from the $SU(\Ncdual) \times 
Sp(2\Ncpdual)$ theory discussed in section~4.   We observe that our original 
dual reduces to a symplectic theory with only fundamental and singlet matter 
fields when the $SU(\Nf)$ gauge coupling is set to zero.  It is then 
straightforward to apply Seiberg's duality to the $Sp(2\Ncpdual)$ gauge group 
\refs{\SeibergI,\IntriligatorPouliot}.  After replacing the $Sp(2\Ncpdual)$ 
theory with its dual and restoring the $SU(\Ncdual)$ gauge coupling, we find 
a new magnetic description of the $SO(10)$ model.  We sketch a derivation 
of this second dual below. 

	We start with the magnetic theory of section~4.  When the
$SU(\Ncdual)$ gauge coupling is turned off, the fields $q'^{\alpha
\adot}_\I$ and $t^\adot_{\dot{\I}}$ in \newmagmatter\ become $2
\Ncdual$ and $2 \NQ-2$ fundamentals under the remaining
$Sp(2\Ncpdual)$ group.
\foot{In this appendix, we adopt the index $\dot{I}$ as shorthand for 
$(I_1 \cdots I_{2N_Q-3})$.}
We recall that the dual to $Sp(2N_c)$ with $N_f$ fundamentals has gauge 
group $Sp(N_f-2N_c-4)$.  The gauge group in our new magnetic theory is thus 
$SU(\Nf+2\NQ-7) \times Sp(2\Nf+4\NQ-18)$. 

	Seiberg's duality transformation introduces certain mesons as 
fundamental fields on the magnetic side. These mesons correspond to 
$Sp(2\Ncpdual)$ invariant combinations of $q'^{\alpha \adot}_{I}$ and 
$t^\adot_{\dot{I}}$:
\eqn\mesons{
\eqalign{ (m^{(a)})^{[\alpha\beta]}_{(\I\J)} &\sim q'^{\alpha \adot}_{(\I}
  q'^{\beta \bdot}_{\J)}~J_{\adot\bdot} \cr
(m_{t})_{[\dot{\I}\dot{\J}]} &\sim t^\adot_{\dot{\I}}t^\bdot_{\dot{\J}}
J_{\adot\bdot} \cr}
\qquad\qquad
\eqalign{
(m^{(s)})^{(\alpha\beta)} &\sim q'^{\alpha \adot}_1
  q'^{\beta \bdot}_2J_{\adot\bdot} \cr
(m_{q't})^\alpha_{\I\dot{\J}} &\sim q'^{\alpha \adot}_\I t^\bdot_{\dot{\J}}
J_{\adot\bdot}.}}
It should be noted that $(m_{q't})^\alpha_{\I\dot{\J}}$ is not irreducible 
under the $SU(2)$ flavor group, but is rather the direct sum of two irreps
with dimensions $2N_Q-1$ and $2N_Q-3$.  When expressed in terms of the mesons, 
the magnetic superpotential of \newWmagnetic\ becomes
\eqn\newsup{\Wmag~=~mqsq+m^{(s)}s+qn\bar{q}+m_{q't}\bar{q}.}
We must also include superpotential couplings between the dual quarks 
$\hat{q}'^{\adot}_{\alpha \I}$ and $\hat{t}^{\adot}_{\dot{\I}}$ and the 
fields in \mesons:
\eqn\dw{\Delta \Wmag~=~m^{(s)}\hat{q}'\hat{q}'+m^{(a)}\hat{q}'\hat{q}'+
            m_{q't}\hat{q}'\hat{t}+m_t \hat{t}\hat{t}.}
The bilinear terms in \newsup\ give mass to $m^{(s)}$, $s$, $\bar{q}$ 
and $2N_Q-1$ components of $m_{q't}$.  Upon integrating out these heavy
fields, we arrive at the new dual theory which has nonabelian symmetry group
\foot{For brevity's sake, we do not display any $U(1)$ charge assignments in 
appendices B, C or D.}
\eqn\symgrouptwo{{\Gdual}' = \bigl[SU(\Nf+2\NQ-7) \times Sp(2\Nf+4\NQ-18) 
\bigr]_{\rm local} \times \bigl[ SU(\Nf) \times SU(2) \bigr]_{\rm global},}
matter content
\eqn\mattertwo{\eqalign{
q & \sim (\fund, 1 ; \antifund,1) \cr
{\tilde q}' &  \sim ( \antifund,\fund; 1, 2) \cr
{\hat t} & \sim (1, \fund; 1, 2 \NQ-2) \cr
m^{(a)} & \sim (\anti, 1; 1,3) \cr
m_{q' t} & \sim (\fund, 1; 1, 2 \NQ-3) \cr
m & \sim (1,1 ; \sym,1) \cr
n & \sim (1,1; \fund, 2\NQ-1) \cr
m_t & \sim (1,1; 1, [(2\NQ-2) \times (2\NQ-2)]_\A ) \cr}}
and tree level superpotential
\eqn\neww{
\Wmag + \Delta \Wmag = mqq\hat{q}'\hat{q}'+qn\hat{q}'\hat{t}
+m^{(a)}\hat{q}'\hat{q}'+m_{q't} {\tilde q}' t + m_{t}\hat{t}\hat{t}.}

	As in the original dual, the Wilsonian beta function coefficients for 
the new magnetic gauge couplings 
\eqn\betanew{\eqalign{
b_0^{SU} &= 12-N_F-2N_Q \cr
b_0^{Sp} &= 2N_F+3N_Q-16}}
indicate the absence of a free magnetic phase.

\appendix{C}{A related duality transformation}

\def\cs{{\hat s}}
\def\cq{{\hat q}}
\def\cqp{{\hat q}'}
\def\cqbar{\bar{\hat q}}
\def\cu{{\hat u}}

\def\cm{{\hat m}}
\def\cn{{\hat n}}
\def\cV{{\cal V}}
\def\cM{{\cal M}}
\def\cN{{\cal N}}

An interesting variant on our duality transformation can be found
by considering a particular relevant perturbation in the form
of a nonzero superpotential. Consider a theory with symmetry group
\eqn\symgroupp{G = SO(10)_\local \times \bigl[ SU(\Nf) \times SU(2)
\bigr]_\glob}
and matter content
\eqn\matterp{\eqalign{
V^i_\mu & \sim \bigl( 10; \fund ,1  \bigr) \cr
Z^X_\mu & \sim \bigl( 10; 1 ,2N_Q-1 \bigr) \cr
Q^\A_\I  & \sim \bigl( 16; 1,N_Q  \bigr) \ .\cr}}
In the absence of a superpotential, this is simply a theory of the type in
\newmatter\ with $\Nf+2\NQ-1$ vectors.  However, using the fact 
that two $N_Q$ representations of $SU(2)$ can be contracted
symmetrically to form a $2N_Q-1$ representation, we can add the
superpotential $\Wtree=Z^XQ_IQ_J$ while preserving the symmetry group
\symgroup.

In the absence of the electric superpotential, the dual representation
is given by an $SU(\Nf+4\NQ-8)\times Sp(2\NQ-2)$ gauge theory.  The
addition of the superpotential can be analyzed using the details of
the duality given in \newmaggroup--\newWmagnetic.  After performing a long 
but straightforward analysis and renaming various fields, one finds a
magnetic theory with symmetry group
\eqn\newmaggroupp{\Gdual = \bigl[SU( \Nf+2\NQ-7) \times Sp(2\NQ-2) 
\bigr]_{\rm local} \times \bigl[ SU(\Nf) \times SU(2) \bigr]_{\rm global}}
and matter fields
\eqn\newmagmatterp{\eqalign{
\cq^\a_i &\sim \bigl( \fund ,1 ; \antifund,1 \bigr) \cr
\cqp^{\a\adot}_\I &\sim \bigl( \fund,\fund; 1,2 \bigr) \cr
\cqbar_{\a (\I_1 \cdots \I_{2\NQ-2})}  &\sim \bigl( \antifund,1; 1,2\NQ-1
  \bigr) \cr
\cs_{\a\b} &\sim \bigl( \symbar,1; 1,1; \bigr) \cr
\hat t^{\adot}_{(\I_1 \cdots \I_{2\NQ-3})} &\sim \bigl( 1,\fund; 1,2\NQ
  \bigr).  \cr
\cm^{(ij)} &\sim \bigl(1,1; \sym,1 \bigr) \cr
\cn^i_{(\I_1 \cdots \I_{2\NQ -2})} &\sim \bigl(1,1; \fund,2\NQ-1 \bigr). \cr}}
The superpotential has the same form as \newWmagnetic\ with the obvious
replacements of fields.  Although this theory looks deceptively
similar to the magnetic theory of Section 4, it is different in important
ways.  First, the field $\hat t$ transforms in the $2N_Q$ dimensional
representation of $SU(2)$, unlike the field $t$ in \newmagmatter\
which is in the $2N_Q-2$.  Furthermore, the ``window-frame'' invariant
$K=Q^4$ is zero because of the superpotential.  Instead the operator
$\hat t^2$ is mapped to $Z^2$.  Also, while the operator $m^{ij}$ is
again the image of $V^iV^j$, the operator $n^i_X$ is now mapped to
$V^iZ_X$.

\appendix{D}{Renormalization group check}

In this appendix, we present the details of the two sets of duality
transitions summarized in \flowone--\flowtwo.  At each stage, we
assume that the gauge groups not involved in the duality transition
are weakly coupled at the energy scale of the transition, so they can
be treated as spectators.  This assumption can always be satisfied for
appropriate choices of the high energy coupling constants.

	At ultrahigh energies, the theory has nonabelian symmetries 
\eqn\sosuspgroup{G = \bigl[SO(2N_Q+N_f-3)\times
SU( \Nf+2\NQ-7) \times Sp(2\NQ-2) \bigr]_{\rm local} \times 
\bigl[ SU(\Nf) \times SU(2) \bigr]_{\rm global}}
and matter fields
\eqn\sosuspmatter{\eqalign{
q &\sim \bigl(1, \fund ,1 ; \antifund,1 \bigr) \cr
\qp &\sim \bigl(1, \fund,\fund; 1,2 \bigr) \cr
\qbar  &\sim \bigl( \antifund,1; 1,2\NQ-1 \bigr) \cr
u &\sim \bigl( \fund,\antifund,1; 1,1 \bigr) \cr
t &\sim \bigl( 1,1,\fund; 1,2\NQ-2 \bigr) \cr 
m &\sim \bigl(1,1,1; \sym,1; \bigr) \cr
n &\sim \bigl(1,1,1; \fund,2\NQ-1 \bigr) \ . \cr}}
These fields are the same as those in \newmagmatter\ except for the
absence of the field $s$ and the presence of a field $u$.  We will soon 
identify $s$ as the bilinear $uu$.  The superpotential for the theory is 
consistent with this identification:
\eqn\Wsosusp{
W = mquuq+\qp uu \qp + nq\qbar + \qp\qbar t\ .}

	Consider the renormalization group flow of \flowone, for which
the $SO$ factor becomes strongly coupled first.  The $SO(\Nf+2\NQ-3)$
gauge group contains $\Nf+2\NQ-7$ vector representations and no other
charged matter.  Such a theory is known to confine and to have an
effective description in terms of a bilinear composite field $uu$
\IntriligatorSeiberg.  No dynamical superpotential is
generated.
\foot{The theory actually has a second branch with a destabilizing dynamical 
superpotential.  We disregard this other branch as no supersymmetric vacua 
are associated with it.} 
The field $uu$, if identified with $s$, leaves the remaining $SU\times Sp$
theory with the gauge group, matter content, and superpotential of the
theory \newmaggroup--\newWmagnetic.  The low energy dynamics is thus
equivalent to $SO(10)$ with $N_f$ vectors, $N_Q$ spinors and 
vanishing superpotential.

	Now let us consider the more intricate behavior of \flowtwo\ which
follows from allowing the gauge coupling of the $SU$ group to grow
strong first.  Since the $SU$ group has only fundamentals and
antifundamentals, we may take its dual using Seiberg's transformation
\SeibergI.  This leads to a low energy $SU(2N_Q+3)$ gauge group.  The
bilinears $A=\qp u$, $B=\qp q$, $C=qu$ and $D=q\qbar$ all become
singlet fields in the low energy theory, while dual quarks $\hat
q,\hat\qp,\hat u, \hat \qbar$ now appear.  The field $B$ is a
reducible representation under the $SU(2)$ flavor group; for reasons
which will become clear we name its two irreducible components $\bar
t$ and $\hat t$.  Under the symmetries of the theory
\eqn\sosusptwogroup{G = \bigl[SO(2N_Q+N_f-3)\times
SU( 2\NQ+3) \times Sp(2\NQ-2) 
\bigr]_{\rm local} \times \bigl[ SU(\Nf) \times SU(2) \bigr]_{\rm global},}
the matter fields transform as
\eqn\dualonematter{\eqalign{
A &\sim \bigl(\fund,1,\fund;1,2\bigr) \cr
\bar t &\sim \bigl(1,1,\fund;1,2N_Q-2\bigr) \cr
\hat t &\sim \bigl(1,1,\fund;1,2N_Q\bigr) \cr
C &\sim \bigl(\fund,1,1;\antifund,1\bigr) \cr
D &\sim \bigl(1,1,1;\antifund,2N_Q-1\bigr) \cr \cr \cr
}
\quad\quad\quad\quad\eqalign{\cq &\sim \bigl( 1,\fund,1;\fund,1\bigr) \cr
\cqp &\sim \bigl( 1,\fund,\fund,1,2\bigr) \cr
\cqbar &\sim \bigl(1,\antifund,1;1,2N_Q-1\bigr) \cr
\cu &\sim \bigl( \fund,\antifund,1;1,1\bigr) \cr
t &\sim \bigl(1,1,\fund;1,2N_Q-2\bigr) \cr
m &\sim \bigl(1,1,1; \sym,1 \bigr) \cr
n &\sim \bigl(1,1,1; \fund,2\NQ-1 \bigr)
\ . \cr } }
The superpotential of this theory is
\eqn\Wsusosptwo{
W = A^2 + \bar t t + A\cqp\cu + \bar t\cqp\cqbar + \hat t\cqp\cqbar 
+ C\cq\cu + D\cq\cqbar + mCC + nD\ .}
The fields $A,\bar t,t,n,D$ are massive and can be integrated out.
This leaves the superpotential
\eqn\Wsosuspthree{
W = \cqp\cu\cu\cqp +  \hat t\cqp\cqbar + C\cq\cu + mCC\ .}

The $SO(N_f+2N_Q-3)$ group can now become strongly coupled.  It has
only vector representations $C$ and $\cu$, so its strong dynamics can
be described using Seiberg's duality
\refs{\SeibergI,\IntriligatorSeiberg}.  The low energy group, independent
of $\Nf$ and $\NQ$, is $SO(10)$.  The fields $C$ and $\cu$ combine
to give $SO(10)$ singlets $E=CC$, $F=C\cu$, and $\cs=\cu\cu$, and
there are dual quarks $V$ and $v$ which are vectors of $SO(10)$.  The
symmetry group is now
\eqn\sosuspthreegroup{G = \bigl[SO(10)\times
SU( 2\NQ+3) \times Sp(2\NQ-2)\bigr]_{\rm local} \times \bigl[ SU(\Nf)
\times SU(2) \bigr]_{\rm global} \ ,}
under which the matter fields transform as
\eqn\dualtwomatter{\eqalign{
m &\sim \bigl(1,1,1;\sym,1 \bigr) \cr
E &\sim \bigl(1,1,1;\symbar,1\bigr) \cr
F &\sim \bigl(1,\antifund, 1; \antifund,1\bigr) \cr
\cq &\sim \bigl( 1,\fund,1;\fund,1\bigr) \cr \cr \cr}
\quad\quad\quad\quad\eqalign{
V &\sim \bigl(\fund,1,1;\fund,1\bigr) \cr
v &\sim \bigl(\fund,\fund, 1;1,1\bigr) \cr
\cqp &\sim \bigl( 1,\fund,\fund,1,2\bigr) \cr
\cqbar &\sim \bigl(1,\antifund,1;1,2N_Q-1\bigr) \cr
\cs &\sim \bigl(1,\symbar,1;1,1\bigr) \cr
\hat t &\sim \bigl(1,1,\fund;1,2N_Q\bigr) 
\ . \cr}}
The superpotential is
\eqn\Wsusospfour{
W = \cqp\cs\cqp + \hat t\cqp\cqbar + \cs v v + F\cq + mE + EVV + FVv \ .}
The fields $m,E,F,\cq$ are massive, and when they are integrated out the
last four terms in the above superpotential are removed.

The $SU\times Sp$ subgroup of this last theory has charged matter
which is similar to the magnetic theory discussed in Appendix~C.
However, there are several minor differences.  First, the field $\cq$
in \newmagmatterp\ is replaced by the field $v$ above.  The fact that
$v$ is charged under $SO(10)$ will not matter as long as $SO(10)$ is
weakly coupled.  Second, the superpotential \Wsusospfour\ contains the
extra term $\cs v v$; this causes no difficulties since it is mapped
under duality to a corresponding term in the electric superpotential.
Third, the mesons $\cm$ and $\cn$ in \newmagmatterp\ are missing,
along with their superpotential couplings.  However, it is easy to
convert the duality of Appendix~C to this situation by adding singlets
$\bar\cm$ and $\bar\cn$ to both sides, along with the superpotential
couplings $\Delta W = \bar\cm\cm + \bar\cn\cn$ on the $SU\times Sp$
side and $\Delta W = \bar\cm VV + \bar\cn VZ$ on the $SO(10)$ side.

Thus, under the duality of Appendix~C, the $SO(10)$ factor is a
spectator, while the $SU\times Sp$ subgroup is transformed into a
second $SO(10)$ factor with $N_Q$ spinors, along with $2N_Q-1$ vectors
$Z^i$ and $10$ fields $\cV$ in the vector representation of both
$SO(10)$ groups.  The fields $\cM=\cs vv$ and $\cN=v\cqbar$ are also present. 
Note that $\cM$ decomposes under the first $SO(10)$ into a
traceless symmetric tensor $\cM_s$ and a singlet $\cM_0$.  The
symmetry group of the theory is
\eqn\sosuspfourgroup{G = \bigl[SO(10)\times
SO(10)\bigr]_{\rm local} \times \bigl[ SU(\Nf)
\times SU(2) \bigr]_{\rm global} \ ,}
and the matter content is
\eqn\sosomatter{\eqalign{
V &\sim \bigl( \fund,1;\fund,1\bigr) \cr
\cV &\sim \bigl(\fund,\fund;1,1\bigr) \cr
Z &\sim \bigl( 1,\fund;1,2N_Q-1\bigr) \cr
Q &\sim \bigl( 1,16;1,N_Q\bigr) \cr}
\quad\quad\quad\quad\eqalign{
\cM_s &\sim \bigl( \sym, 1; 1, 1\bigr) \cr
\cM_0 &\sim \bigl( 1 ,1;1,1 \bigr) \cr
\cN &\sim \bigl( \fund,1;1,2N_Q-1\bigr) \ . \cr \cr
}}
The superpotential of the theory is
\eqn\sososup{
W = ZQQ + \cM_0 + (\cM_s+\cM_0)\cV\cV + \cN\cV Z \ .}
The linear term $\cM_0$ is the image under duality of the
extra term $\cs v v$ in the superpotential \Wsusospfour.

Because of the linear term, the equation of motion for $\cM_0$ causes
tr$\cV\cV$ to condense.  D-term conditions force the field $\cV$ to develop a
diagonal expectation value which breaks $SO(10)\times SO(10)$ to the
diagonal $SO(10)$ subgroup.  The $\cN\cV Z$ term gives mass to $Z$ and
$\cN$, while the $\cM\cV\cV$ term gives mass to $\cM_s$, $\cM_0$ and
the 55 components of $\cV$ which are not eaten by gauge bosons.  This
eliminates the entire superpotential.  The remaining fields are $N_f$
vectors $V$ and $N_Q$ spinors $Q$ under the $SO(10)$ gauge group.
This is the endpoint of the flow \flowtwo, and it matches with the
endpoint of \flowone.

\listfigs
\listrefs
\bye